\documentclass{aa}
\usepackage{epsfig}
\begin{document}

   \thesaurus{11     % A&A Section 3: Extragalactic Astronomy
              (11.01.1;  % Galaxies: abundances
               11.05.1;  % Galaxies: elliptical and lenticular, cD
               11.05.2;  % Galaxies: evolution
               11.09.4;  % Galaxies: ISM
               09.16.1)}  % (ISM:) planetary nebulae: general
   \title{Planetary nebulae in \object{M32} and the bulge of \object{M31}:\\
   Line intensities and oxygen abundances}

%   \subtitle{}

   \author{Michael G. Richer \inst{1}
          \thanks{
           The overwhelming majority of my work on this project
           was done while I was at the Observatoire de Meudon as
           a member of the DAEC.}
           \fnmsep
           \thanks{
           Visiting Astronomer, Canada-France-Hawaii Telescope, operated by the
           National Research Council of Canada, the Centre National de la Recherche
           Scientifique de France, and the University of Hawaii}
          \and Gra\.zyna Stasi\'nska \inst{2}$\ {}^{\star \star }$
          \and Marshall L. McCall \inst{3}}

   \offprints{M. G. Richer}

   \institute{Instituto de Astronom\'{\i}a, UNAM,
              Apartado Postal 70-264, 04510 M\'exico, D. F., M\'exico\\
              email: richer@astroscu.unam.mx
              \and
              DAEC, Observatoire de Meudon,
              5 Place Jules Janssen, F-92195 Meudon Cedex, France\\
              email: grazyna@obspm.fr
              \and
              Dept. of Physics and Astronomy, York University,
              4700 Keele Street, Toronto, Ontario, Canada   M3J 1P3\\
              email: mccall@aries.phys.yorku.ca}

   \date{Received ? / Accepted ?}
   \titlerunning{Spectroscopy of planetary nebulae in M32 and the bulge of M31}
   \maketitle

\begin{abstract}

We  present spectroscopy of planetary nebulae in \object{M32} and
in   the  bulge  of  \object{M31}  that  we  obtained  with  the   MOS
spectrograph  at  the  Canada-France-Hawaii  Telescope.   Our
sample  includes 30 planetary nebulae in \object{M31} and 9  planetary
nebulae in \object{M32}.  We also observed one
\ion{H}{ii} region in the disk
of  \object{M31}.   We  detected [\ion{O}{iii}]$\lambda$4363 in 18  of
the  planetary
nebulae,  4  in \object{M32} and 14 in the bulge of \object{M31}.
We  use  our
line  intensities to derive electron temperatures and  oxygen
abundances for the planetary nebulae.


\end{abstract}

%
%________________________________________________________________

\section{Introduction}

One  of  the most important clues concerning  the  early evolution
of  dynamically  hot galaxies  (DHGs:  ellipticals, dwarf
spheroidals, and bulges of spirals) in the fundamental plane   of
galaxies  is the  existence  of  a  well-defined relationship
between metallicity and  mass  (e.g., Bender et al.
\cite{Benderetal1993}). The fundamental lesson  taught  by this
relation is that star formation in DHGs stopped because of  gas
loss, with less massive systems losing greater  fractions of their
gas.  Outflow probably begins when supernovae have raised the
internal energy of the gas enough to allow it to escape the
potential well (e.g., Brocato et al. \cite{Brocatoetal1990}).

Most commonly, the metallicity in DHGs is measured via the Mg$_2$
index. While the Mg$_2$ index is an excellent means of ranking
galaxy metallicities, it does not yield an abundance directly,
i.e., the number density of a particular element relative to
hydrogen, and calibrations of the Mg$_2$ index (model-dependent)
are usually in terms of the iron abundance, an element whose
production is notoriously difficult to model.  Though this may be
best for some purposes, e.g., studies of stellar populations, it
is not sufficient for all purposes.  To study the chemical
evolution of DHGs requires the abundance of an element whose
production is well understood. Were such abundances available,
there would be some hope of quantifying the gas fraction at which
DHGs of different masses begin to lose mass.  Knowledge of the
abundances would admit studying the yield of heavy elements, and
hence the slope of the stellar initial mass function during the
star formation epoch. Given the known photometric and dynamical
properties of DHGs today, abundances would also allow us to study
the global energetics involved during their star formation phase.

This  paper  is one of a sequence investigating  the oxygen
abundances of DHGs. Here,  we   present   oxygen abundances for
samples of planetary nebulae in \object{M32} and in the bulge of
\object{M31}. These two  nearby  systems   are   good
representatives of typical DHGs. Though \object{M32}'s light
profile may  be truncated  compared to  isolated  ellipticals, its
structural, dynamical, and spectral properties are  perfectly
typical  for an elliptical of its luminosity (Kormendy
\cite{Kormendy1985}; Bender et al. \cite{Benderetal1993}).
Similarly, recent  work  on  the DHG fundamental plane has shown
that the photometric,  dynamical, and stellar population
properties of bulges follow those  of pure ellipticals (Bender et
al. \cite{Benderetal1992}, \cite{Benderetal1993}).

\begin{table*}
\caption[]{Observing Log}
\label{table1}
\[
\begin{tabular}{lllllll}
\hline
\noalign{\smallskip}
Date  &  Object & Grism & Dispersion & Spectral Range$^a$ & Exposures & Standards \\
\noalign{\smallskip}
\hline
\noalign{\smallskip}
1994 Aug 1/2 & \object{M32}      & U900 & 0.9\AA/pixel & 3727\AA-H$\gamma$
& $1 \times 2700$s & \object{Feige 15} \\
1994 Aug 2/3 & \object{NGC 6720} & U900 & 0.9\AA/pixel & 3600\AA-5400\AA
& $3 \times 1500$s & \object{Feige 15} \\
             &          &      &              &
& $1 \times 1200$s &          \\
             & \object{M32}      & U900 & 0.9\AA/pixel & 3727\AA-H$\gamma$
& $3 \times 2700$s & \object{Feige 15} \\
             & \object{M32}      & O300 & 3.5\AA/pixel & 4686\AA-1$\mu$m
& $1 \times 900$s  & \object{Wolf 1346} \\
             &          &      &              &
& $1 \times 1800$s &           \\
1994 Aug 3/4 & \object{M32}      & B600 & 1.5\AA/pixel & 3727\AA-5876\AA
& $1 \times 1800$s & \object{Wolf 1346} \\
             &          &      &              &
&                  & \object{Feige 15} \\
             & \object{M31}      & B600 & 1.5\AA/pixel & 3727\AA-5876\AA
& $3 \times 2700$s & \object{Wolf 1346} \\
             &          &      &              &
& $1 \times 1200$s & \object{Feige 15} \\
\noalign{\smallskip}
\hline
\end{tabular}
\]
\begin{list}{}{}
\item[$^{\mathrm{a}}$] These are {\it minimum} spectral ranges.  The
actual spectral range will depend upon the object's position
within the spectrograph's field of view.
\end{list}
\end{table*}

Oxygen  is an excellent element with which to study  the
evolution  of  galaxies.  Oxygen is a primary  element  whose
sole  significant  production  site  is  type  II  supernovae
(Wheeler et al.
\cite{Wheeleretal1989}),
so its abundance  is  tied
directly  to the history of massive star formation,  and  the
enrichment   time  scale  is  short  compared  to   the   gas
consumption  time scale.  Oxygen abundances are  also  easily
observable in planetary nebulae. Planetary nebulae have  high
electron    temperatures,   so   the    temperature-sensitive
[\ion{O}{iii}]$\lambda$4363  line  is  observable,  making  it
possible   to
determine  accurate electron temperatures in high metallicity
environments.   Further, the dominant  ionization  stages  of
oxygen,  O${}^{+}$  and  O${}^{++}$,  have observable  lines,
while  other
ionization  stages are easily accounted for using  ratios  of
readily  detectable helium lines (e.g., Kingsburgh  \&  Barlow
\cite{KingsburghBarlow1994}).

Planetary nebulae are good sites in which to probe the oxygen
abundance, and  they are the  only sites that are directly
accessible in DHGs.  Since planetary nebulae are bright in strong
emission lines (e.g., [\ion{O}{iii}]$\lambda$5007), they are
easily located within their parent galaxies using emission-line
and continuum-band imaging (e.g., Ciardullo et al.
\cite{Ciardulloetal1989}). Observational and theoretical evidence
indicates  that  the stellar  precursors of most planetary nebulae
do  not  modify their  initial oxygen abundance (Iben \& Renzini
\cite{IbenRenzini1983}; Henry \cite{Henry1989}; Perinotto
\cite{Perinotto1991}; Forestini \& Charbonnel
\cite{ForestiniCharbonnel1997}). Hence, a planetary  nebula's
oxygen abundance reflects  that  in  the interstellar medium at
the time of its precursor's formation.  Finally, most of the
stellar populations  in  DHGs are  old, so they will produce
planetary nebulae at comparable rates per unit mass.  As a result,
planetary nebulae sample the oxygen abundances in DHGs according
to the mass in each stellar population.  The resulting mean oxygen
abundance for the planetary nebula population in a DHG should then
be a mass-weighted mean of the oxygen abundances in its stellar
populations.

Apart from their utility for studying the chemical evolution of
\object{M31} and \object{M32}, the spectroscopic data for the
planetary  nebulae we present are interesting for what they reveal
about the evolution of the planetary nebulae themselves. Though
there may exist a good qualitative understanding of planetary
nebula evolution, it is unclear how well it stands up to
quantitative scrutiny.  This situation arises primarily because
the distances to planetary nebulae  are difficult to establish
within  the \object{Milky Way}. Traditionally, this constraint has
made it difficult to study such absolute properties as the
luminosity and size of planetary nebulae, as well as the temporal
evolution of these quantities. Extragalactic planetary nebulae are
especially valuable  in this regard because their distances  are
known. The addition of the data sets for \object{M32} and the
bulge of \object{M31} is particularly helpful since these
planetary nebulae arise from old stellar  populations. They  will
thus   provide   an intriguing contrast with the planetary nebula
populations  in the \object{Magellanic Clouds}, which are the
product of recent star formation  (Richer \cite{Richer1993}).
Whether the evolution of planetary nebulae depends upon the
progenitor mass or metallicity are among  the questions that we
may hope to  answer  through  a comparison of the properties of
planetary nebulae in \object{M31} and \object{M32}    with   those
elsewhere.    A   better quantitative understanding of planetary
nebula evolution would be a  great help   in  understanding  and
using  the  planetary nebula luminosity function as a distance
indicator.

In this paper, we present our spectroscopic data for our
samples of planetary nebulae in \object{M32} and in the bulge of
\object{M31}.
The  observations  and  their  reductions  are  described  in
Section 2.  The line intensities and reddenings we deduce are
presented   in  Section  3.   The  reddening-corrected   line
intensities  are then used to calculate electron temperatures
and  oxygen  abundances in Section 4.  Summary  comments  are
given in Section 5.

In companion papers, we will use the  data  we present below to
study the chemical evolution of DHGs and the evolution of
planetary nebulae in different environments.

\section{Observations and reductions}

\begin{table*}
\caption[]{B600/O300 Sensitivity Correction} \label{table2}
\[
\begin{tabular}{lll}
\hline \noalign{\smallskip} Wavelength                 &
I(B600)/I(O300) & \object{M32} Planetary Nebulae Used${}^{\mathrm
a}$ \\ \noalign{\smallskip} \hline \noalign{\smallskip} H$\beta$
& $1.17 \pm 0.06$   & all, except PN25 \\
$[\ion{O}{iii}]\lambda$4959       & $1.00 \pm 0.01$   & PN1, PN2, PN4, PN5, PN7, PN8, PN17, %%@
PN24 \\ \ion{He}{i}\,$\lambda$5876  & $0.77 \pm 0.03$   & PN1,
PN7, PN8, PN11, PN17 \\ H$\alpha$                  & $0.85 \pm
0.02$   & PN2, PN8, PN11, PN17, PN24 \\ \noalign{\smallskip}
\hline
\end{tabular}
\]
\begin{list}{}{}
\item[${}^{\mathrm a}$] PN4 and PN17 are background objects in the disk
of \object{M31} (Ford \& Jenner \cite{FordJenner1975}).
\end{list}
\end{table*}

Our  ultimate purpose for making these observations  was to  study
the chemical evolution of \object{M32} and the bulge of
\object{M31}. In  \object{M31},  we  chose planetary nebulae in
the inner bulge  in order  to probe the highest levels of
enrichment.  On account of the bright galaxy background, we also
preferentially chose planetary   nebulae  that  were  known  to
be   bright   in [\ion{O}{iii}]$\lambda$5007.  All of the objects
we observed in \object{M31} are found within  the inner half
effective radius of \object{M31}'s  bulge.   In \object{M32},   we
observed  as  many  objects  as  we  could,  again emphasizing
bright  objects  on  account  of   the   galaxy background.   In
this case, the objects we observed  extended to many effective
radii.

We obtained our observations over three nights in August 1994  at
the Canada-France-Hawaii Telescope (CFHT) with  the multi-object
spectrograph (MOS).  The  MOS  is  an  imaging, multi-slit
spectrograph  that  employs  a  grism   as   the dispersing
element  (see Le F\`evre et al. \cite{LeFevreetal1994} for
details). Objects are selected for spectroscopy using focal plane
masks that are constructed on-line from previously acquired
images. The detector was the Loral3 CCD, a thick CCD with 15$\mu$m
square pixels in a $2048\times 2048$ format, coated to enhance the
quantum efficiency in the blue. The Loral3's read noise was  8
electrons and its gain was set to 1.9\,electrons/ADU. For the
observations of both \object{M31} and \object{M32}, we used slits
15\arcsec\ long by 1\arcsec\ wide.  No order-sorting filter was
used for any of these observations.

Table \ref{table1} presents a log of our observations.
During  the
course of the observations, we used three different grism
set-ups in order to optimize throughput, wavelength coverage, and
spectral resolution.  We used the B600 grism only because  of
the disappointing throughput of the U900 grism.  Although the
precise  dispersion and wavelength coverage depend upon  each
object's  position within the field of view,  Table
\ref{table1}  lists
typical  values for all three grisms (minimal ranges for  the
wavelength coverage).

We  used  the standard IRAF routines to reduce the  data
(noao.imred.ccdred),  and  followed  the  standard  reduction
procedure.  First, the overscan bias was removed from all  of
the  images.   Next, for the first two nights,  sequences  of
zero  exposure images were combined and subtracted  from  the
other  images to remove any bias pattern.  This was not  done
on  the  third night because the CCD dewar began to  warm  up
before  we  had a chance to obtain the zero exposure  images.
This  is  unlikely to be a limitation, since no bias  pattern
was  obvious  on  either of the first two  nights.   Finally,
pixel-to-pixel variations were removed using spectra  of  the
internal quartz lamp.

Extracting the spectra proved challenging on account  of
the  nature  and faintness of the sources, and on account  of
the  characteristics  of  the  spectrograph.   The  planetary
nebulae  in \object{M31} and \object{M32} are sufficiently
faint that  we  were
unable to detect their continuum emission.  Only the emission
lines  were visible, appearing as a sequence of dots,  so  it
was  impossible  to  trace these spectra.   Furthermore,  the
spectra  spanned  the  full width of the  detector,  so  they
suffered  from geometric distortion (pin-cushion)  introduced
by  the  optics of the spectrograph.  Fortunately, we had  to
include star apertures when defining the spectrograph's focal
plane mask to permit accurate re-alignment on the field  when
ready to do spectroscopy.  We used these stars (6 for
\object{M31},  3
for  \object{M32})  to  map the geometric distortion
imposed  by  the
optics, and corrected this distortion using the tasks in  the
noao.twodspec.longslit  package  (Anderson
\cite{Anderson1987}).
At  this
point,  we  had  images  in  which the  wavelength  axis  was
parallel  to  the  rows  of the CCD, and  we  could  use  the
brightest  line  in  each spectrum to  define  an  extraction
aperture  (e.g., Massey et al.
\cite{Masseyetal1992}).
Except  for
the  U900 spectra, the individual spectra were extracted from
each  image  and then combined to produce the final  combined
spectra.  To better define the extraction apertures  for  the
U900   spectra,  the  spectra  were  combined  first,   after
verifying that the individual images had the same spatial
coordinate  scales.   In all cases, extraction  involved  local
subtraction of the underlying galaxy and sky spectra.

Establishing a consistent sensitivity scale  across  all three
grism set-ups was a primary consideration of our  data reduction.
We  calibrated the instrumental sensitivity  for each  set-up
using  observations of  the  spectrophotometric standard  stars
listed in Table \ref{table1}.  We verified that our
slitlet-to-slitlet sensitivity scale was secure in three ways.
First, the observations of the standard stars were made in pairs
through two different slitlets.  These slitlets were cut at the
red and blue extremes of the field of view to ensure that our
standard star observations spanned the full wavelength range of
our planetary nebula observations.  These paired observations of
the standard stars had 500\AA, 800\AA, and 1900\AA\ of spectrum in
common for the U900, B600, and O300 grisms, respectively.  In
these overlap regions, the sensitivity functions for each grism
(on each night) were in agreement. Second, we obtained a
spectroscopic sky flat through the standard star mask with the
B600 grism on the last night. This mask contained two slitlets in
addition to those used for the standard star observations.
Comparing the night sky spectra through these four slitlets
indicates that variations in the wavelength sensitivity between
different slitlets are less than 4.5\% (rms). Finally,
observations of NGC 6720 were obtained through a different mask
than the standard stars, and no wavelength-dependent trends are
seen in its sensitivity calibration (see Table \ref{table3}
below).  Therefore, though we did not observe the standard stars
through the slitlets used for our program objects, we have no
reason to believe that our sensitivity calibration is
slitlet-dependent.

We then chose  the  O300 observations of the planetary nebulae in
\object{M32} as our reference data  set.   This  choice  was
motivated  by  a  number of considerations.  First, these
planetary nebulae were observed with  all  three  grisms.  Second,
the O300  grism  has  good sensitivity  over the H$\beta$ --
H$\alpha$ wavelength range (Le F\`evre  et al.
\cite{LeFevreetal1994}), which  contains the strongest lines in
the  spectra. Third, our reddening values for these planetary
nebulae  (see Tables \ref{table4}, \ref{table4b}, and
\ref{table4c}) were  reasonable,  typically $E(B-V)<0.2$\,mag, and
invariably  positive.  These reddenings were consistent  with
previous  observations of PN1 in \object{M32} (Ford et al.
\cite{Fordetal1978}). The  reddening towards \object{M32} is also
expected  to  be small if it is in front of the disk of
\object{M31} (e.g., Burstein  \& Heiles
\cite{BursteinHeiles1984}).

We  ensured  that  there were no systematic  differences between
the  B600  and  O300  data  sets  by  comparing  the intensities
of  H$\beta$, H$\alpha$, [\ion{O}{iii}]$\lambda$4959, and
\ion{He}{i}\,$\lambda$5876  measured relative to
[\ion{O}{iii}]$\lambda$5007 for the planetary nebulae in
\object{M32}.  In making  these  comparisons, we considered only
those  objects for  which  we had the best detections of these
lines.   For these objects, we computed the ratio of the line
intensity in the B600 spectrum to that in the O300 spectrum. Table
\ref{table2} lists the mean value of this ratio, the standard
error in the mean, and  the  objects we considered for each line.
Clearly,  the main  wavelength-dependent trend in Table
\ref{table2} is  a  systematic decrease  in  the  B600 sensitivity
relative  to  the  O300 sensitivity  as  one  goes to  longer
wavelengths.   Simply fitting  a  line to the values in Table
\ref{table2} as  a function  of wavelength, however, yields a
rather poor correction  at  H$\alpha$. As  a result, for
wavelengths between any two lines found  in Table \ref{table2}, we
corrected  for the difference  in sensitivity calibrations by
interpolating   linearly   between    the corrections in Table
\ref{table2}.  For lines to the blue of H$\beta$  or  to the red
of  H$\alpha$,  we  adopted  the  H$\beta$  or  H$\alpha$
corrections, respectively.  We wondered if the upturn at H$\alpha$
in Table \ref{table2} could be due to second order contamination,
but this seems unlikely. Both the O300 and B600 grisms have very
low efficiency at 3250\AA, and a second order contamination would
affect the sensitivity calibration for both grisms similarly.
Consequently, the upturn at H$\alpha$ appears to be real.   The
corrections in Table \ref{table2} were applied to the spectra of
the planetary nebulae in both \object{M32} and the bulge of
\object{M31}.

\begin{table}
\caption[]{Hydrogen Line Intensities for \object{NGC 6720}}
\label{table3}
\[
\begin{tabular}{llrl}
\hline
\noalign{\smallskip}
           & Line      & Intensity${}^{\mathrm{a}}$ & $E(B-V)^{\mathrm{a}}$ \\
\noalign{\smallskip}
\hline
\noalign{\smallskip}
Aperture 1 & H$\beta$   & $100.00 \pm 1.49$ &       \\
           & H$\gamma$  & $42.07 \pm 0.55$ & $0.287 \pm 0.034$ \\
           & H$\delta$  & $24.03 \pm 0.33$ & $0.133 \pm 0.023$ \\
           & H$\epsilon$& $12.99 \pm 0.87$ & $0.279 \pm 0.093$ \\
           & H$8$       & $ 7.14 \pm 0.35$ & $0.484 \pm 0.061$ \\
           & H$9$       & $ 5.79 \pm 0.13$ & $0.280 \pm 0.027$ \\
           & H$10$      & $ 3.97 \pm 0.13$ & $0.331 \pm 0.037$ \\
           & H$11$      & $ 2.85 \pm 0.16$ & $0.371 \pm 0.060$ \\
           & H$12$      & $ 2.32 \pm 0.15$ & $0.304 \pm 0.070$ \\
Aperture 2 & H$\beta$   & $100.00 \pm 1.53$ &       \\
           & H$\gamma$  & $42.29 \pm 0.57$ & $0.273 \pm 0.035$ \\
           & H$\delta$  & $24.06 \pm 0.34$ & $0.131 \pm 0.024$ \\
           & H$\epsilon$& $13.13 \pm 0.90$ & $0.264 \pm 0.094$ \\
           & H$8$       & $ 7.24 \pm 0.36$ & $0.467 \pm 0.063$ \\
           & H$9$       & $ 5.59 \pm 0.19$ & $0.320 \pm 0.041$ \\
           & H$10$      & $ 3.89 \pm 0.15$ & $0.355 \pm 0.043$ \\
           & H$11$      & $ 2.84 \pm 0.19$ & $0.375 \pm 0.073$ \\
           & H$12$      & $ 2.19 \pm 0.18$ & $0.366 \pm 0.089$ \\
Aperture 3 & H$\beta$   & $100.00 \pm 1.45$ & \\
           & H$\gamma$  & $42.51 \pm 0.56$ & $0.259 \pm 0.035$ \\
           & H$\delta$  & $24.72 \pm 0.37$ & $0.085 \pm 0.025$ \\
           & H$\epsilon$& $13.11 \pm 0.84$ & $0.265 \pm 0.088$ \\
           & H$8$       & $ 7.91 \pm 0.43$ & $0.355 \pm 0.068$ \\
           & H$9$       & $ 5.58 \pm 0.26$ & $0.323 \pm 0.056$ \\
           & H$10$      & $ 3.77 \pm 0.22$ & $0.391 \pm 0.064$ \\
           & H$11$      & $ 3.15 \pm 0.37$ & $0.262 \pm 0.129$ \\
           & H$12$      & $ 2.60 \pm 0.37$ & $0.184 \pm 0.151$ \\
\noalign{\smallskip}
\hline
\end{tabular}
\]
\begin{list}{}{}
\item[$^{\mathrm{a}}$] The derivation of the uncertainties in the line
intensities and reddening is described in Sec. 3.
\end{list}
\end{table}

The  U900 data required no correction to put them on the
O300   sensitivity  scale.   We  deduced  this  from   direct
comparison  with  the  B600  and O300  data  (Tables \ref{table4},
\ref{table4b}, and \ref{table4c}),  and
independently  using a spectrum we obtained of  the  Galactic
planetary nebula \object{NGC 6720}. Table \ref{table3} lists the
intensities  and
reddening values for hydrogen lines in three regions  of  NGC
6720.   The reddening values we derive from H$\gamma$, H$\epsilon$,
H$9$,  H$10$,
H$11$, and H$12$ are  in  very good  agreement  in  all  three
apertures,  indicating that our U900 sensitivity  calibration
is   good  to  3750\AA.   Our  reddening  values  at H$\delta$ are
consistently 0.16\,mag lower than calculated from  H$\gamma$,  so  our
U900  sensitivities may be under-estimated by 15\% near 4100\AA.
Our H$8$ reddening values are consistently high, but H$8$ was
blended  with  He~{\sc i}\,$\lambda$3889.   We corrected  the
blend  for  the
He~{\sc i}\,$\lambda$3889  contribution using the
He~{\sc i}\,$\lambda$4471 intensity  assuming
no   radiative  transfer  correction,  thereby  removing  the
maximum  possible He~{\sc i}\,$\lambda$3889 contribution (e.g.,  Aller
\cite{Aller1987}).
Thus, it is perhaps not surprising that our H$8$ reddenings are
too  high.  Overall, our Balmer line intensities for \object{NGC 6720}
indicate that our U900 sensitivity calibration is secure from
3750\AA\  to  H$\beta$.  Similarly, for the planetary nebulae  in
\object{M32}
(Tables \ref{table4}, \ref{table4b}, and \ref{table4c}),
the  U900  line  intensities  for  [\ion{O}{ii}]$\lambda$3727,
[\ion{Ne}{iii}]$\lambda$3869,  and
\ion{He}{ii}\,$\lambda$4686 are in excellent agreement  with
their B600 and O300 counterparts.

\begin{figure*}
%\resizebox{\hsize}{!}{\includegraphics{oap1to6.eps}}
\vspace{-0.2in}
%\hbox{
%\hspace{0cm}\psfig{figure=oap1to6.eps,width=16.5cm,height=10.0cm,angle=90}
\epsfig{file=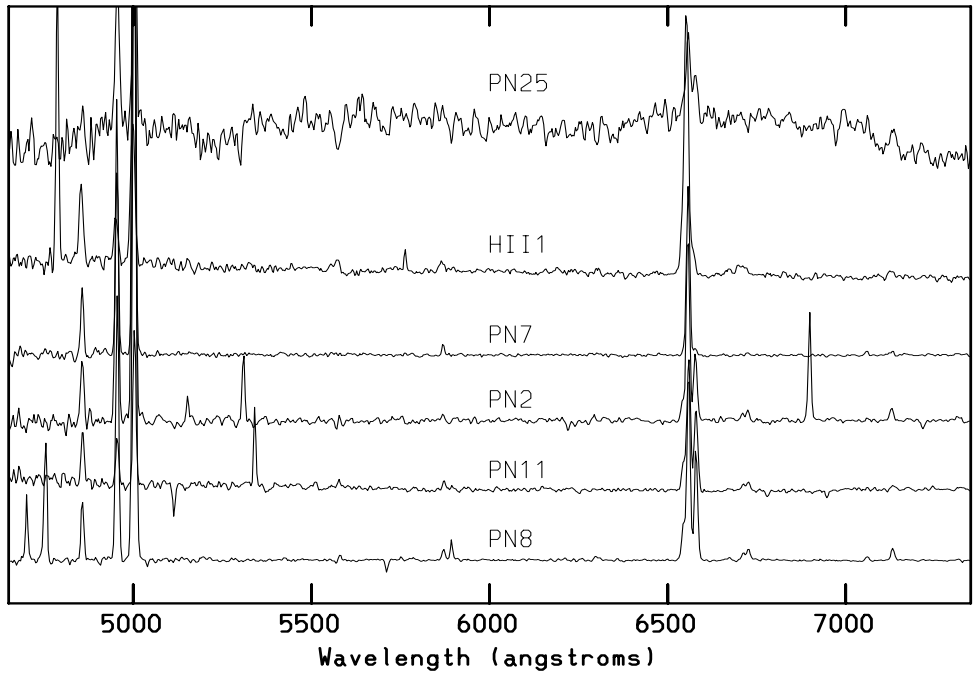,width=10.0cm, height=16.5cm,angle=90,clip}
%}
\caption{
The O300 spectra for PN8, PN11, PN2, PN7,
\ion{H}{ii}\,1,
and  PN25  in the \object{M32} field.  The spectra are displayed  such
that H$\beta$ spans the entire free intensity scale.  Consequently,
lines stronger than H$\beta$ overlap in adjacent spectra.  We  show
only  the  spectral range blueward of 7350\AA.  In all  of  the
spectra we present, cosmic rays were not removed unless  they
interfered with measuring line intensities, so many obviously
remain.   PN25  is very close to \object{M32}'s nucleus,  so  the  sky
subtraction is poorer for this object.  \ion{H}{ii}\,1 is an
\ion{H}{ii} region
in the background disk of \object{M31} (Ford \& Jenner
\cite{FordJenner1975}).
}
\label{fig1a}
\end{figure*}

\begin{figure*}
%\resizebox{\hsize}{!}{\includegraphics{oap7to12.eps}}
\vspace{-0.2cm}
\hbox{
\hspace{0cm}\epsfig{figure=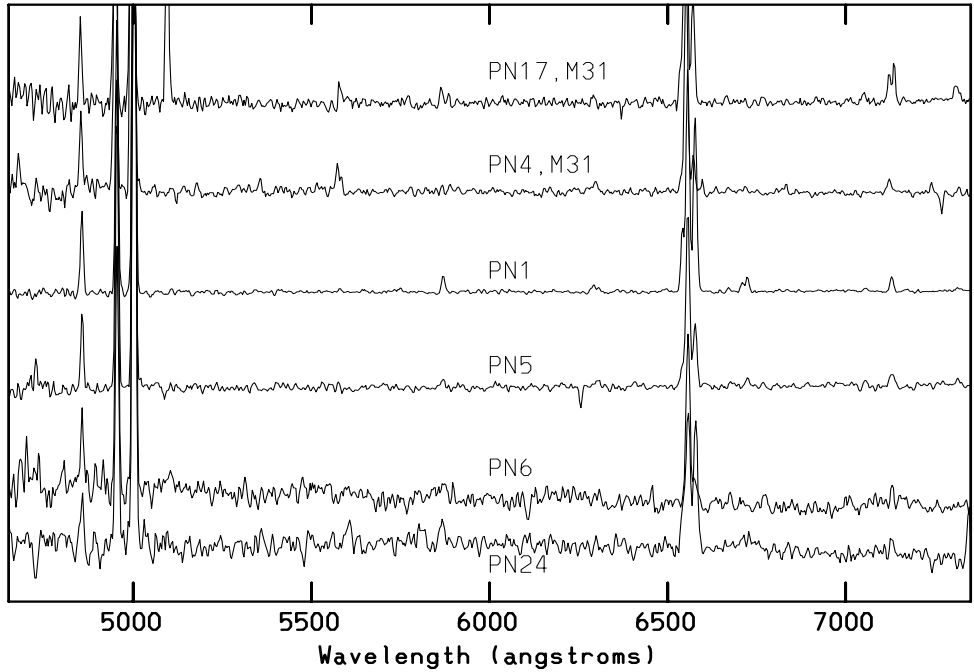,height=16.5cm,width=10.0cm,angle=90}
}
\caption{
The O300 spectra for PN24, PN6, PN5,  PN1,  PN4,
and PN17 in the \object{M32} field.  The format is identical to Fig. \ref{fig1a}.
Like PN25, PN24 is also very close to \object{M32}'s nucleus and
suffers  from  somewhat poorer background subtraction.   Note
that  PN4  and PN17 are background planetary nebulae  in  the
disk of \object{M31} (Ford \& Jenner
\cite{FordJenner1975}).
}
\label{fig1b}
\end{figure*}

\begin{figure*}
%\resizebox{\hsize}{!}{\includegraphics{bap1to6.eps}}
\vspace{-0.5cm}
\hbox{
\epsfig{figure=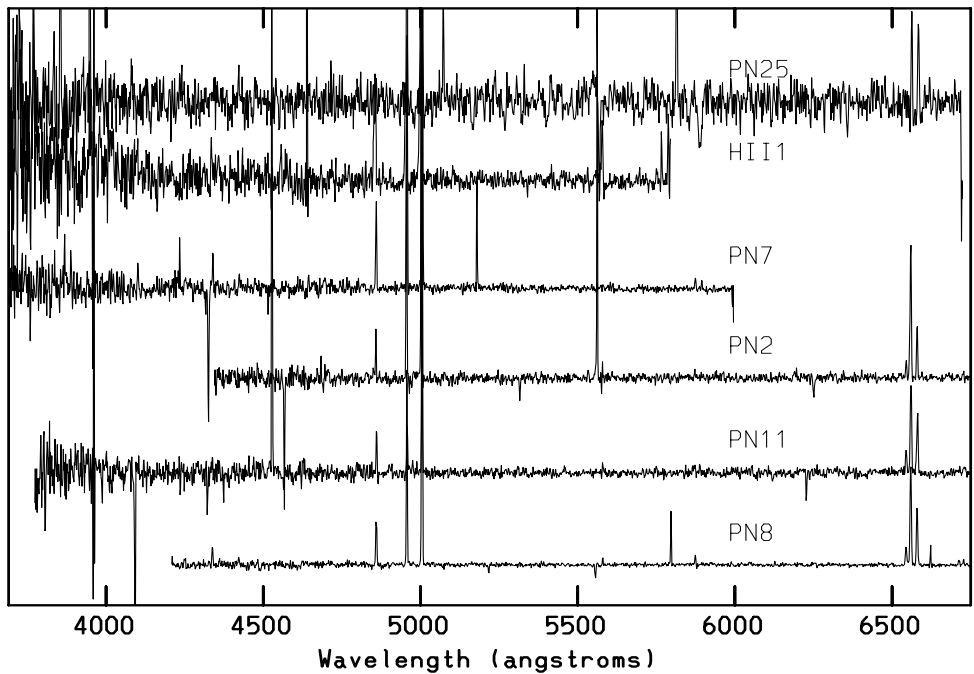,height=16.5cm,width=10.0cm,angle=90}
}
\caption{
The B600 spectra for PN8, PN11, PN2, PN7,  \ion{H}{ii}~1,
and  PN25  in  the \object{M32} field.  For PN8, PN11, and  PN25,  the
scaling  is such that H$\alpha$, not H$\beta$, defines the free  intensity
range.   The  full  useful wavelength range of  the  spectra,
3690\AA\ to  6750\AA,  is shown.  See Fig. \ref{fig1a} for  comments  on
individual objects.
}
\label{fig2a}
\end{figure*}

\begin{figure*}
%\resizebox{\hsize}{!}{\includegraphics{bap7to12.eps}}
\epsfig{figure=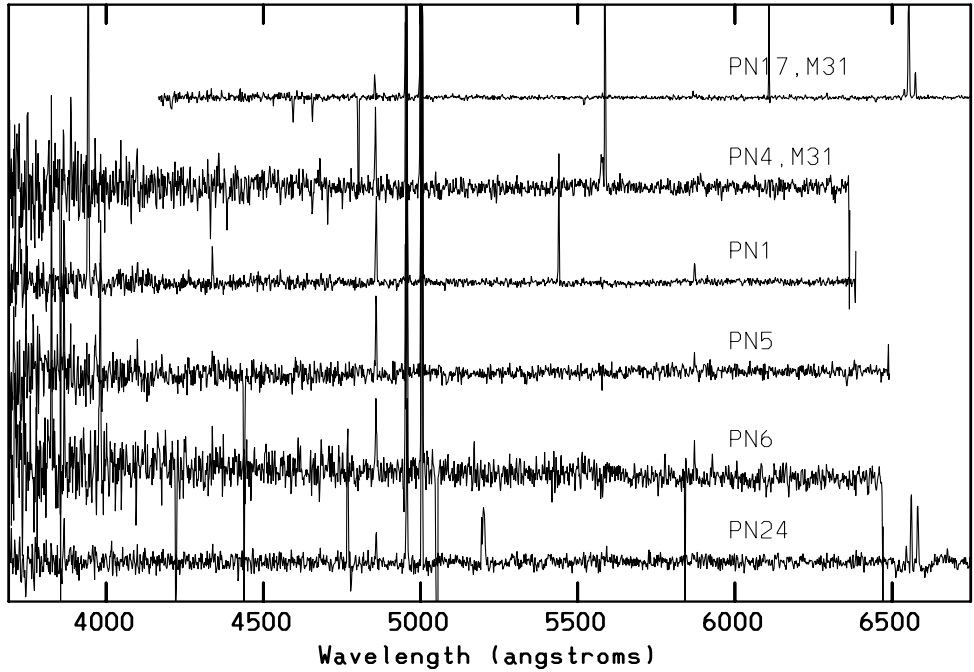,height=16.5cm,width=10.0cm,angle=90}
\caption{
The B600 spectra for PN24, PN6, PN5,  PN1,  PN4,
and  PN17 in the \object{M32} field.  For PN24 and PN17 (\object{M31}),
H$\alpha$, and
not H$\beta$,  defines the free intensity range.  The full  useful
wavelength  range of the spectra, 3690\AA\ to 6750\AA,  is  shown.
See Fig.~\ref{fig1b} for comments on individual objects.
}
\label{fig2b}
\end{figure*}

\begin{figure*}
%\resizebox{\hsize}{!}{\includegraphics{uap1to6.eps}}
\vspace{-0.5cm}
\hbox{
\epsfig{figure=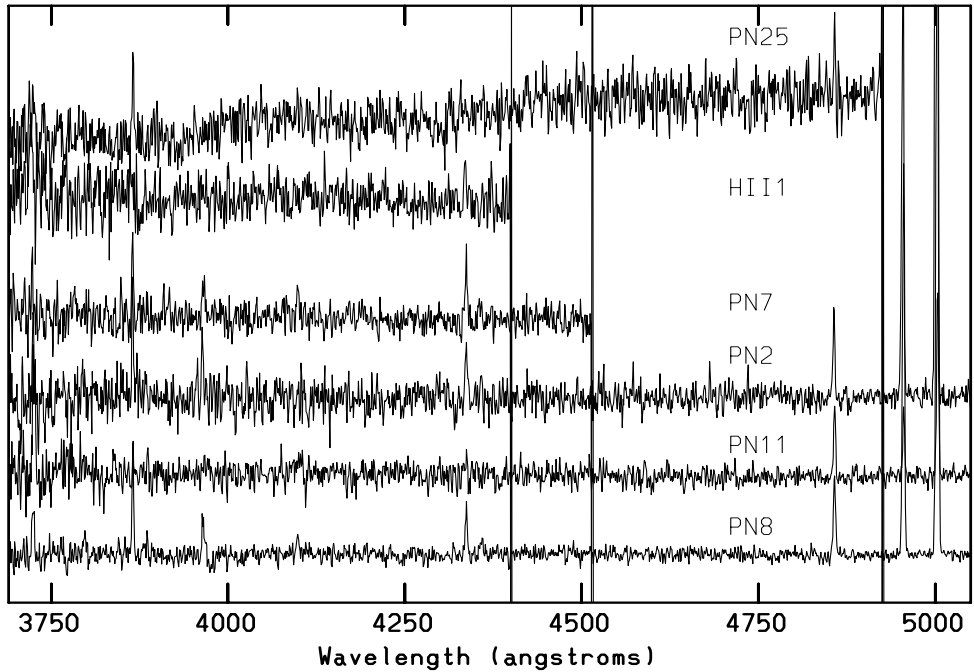,height=16.5cm,width=10.0cm,angle=90}
}
\caption{
The U900 spectra for PN8, PN11, PN2, PN7,  \ion{H}{ii}~1,
and  PN25 in the \object{M32} field.  Only for PN11 and PN25  does  H$\beta$
define the full intensity scale.  For the other objects,  the
full  intensity  scale is defined by H$\gamma$.   The  full  useful
wavelength  range, 3690\AA\ to 5050\AA, is displayed.  See
Fig.~\ref{fig1a} for comments on individual objects.
}
\label{fig3a}
\end{figure*}

\begin{figure*}
%\resizebox{\hsize}{!}{\includegraphics{uap7to12.eps}}
\vspace{-0.5cm}
\hbox{
\epsfig{figure=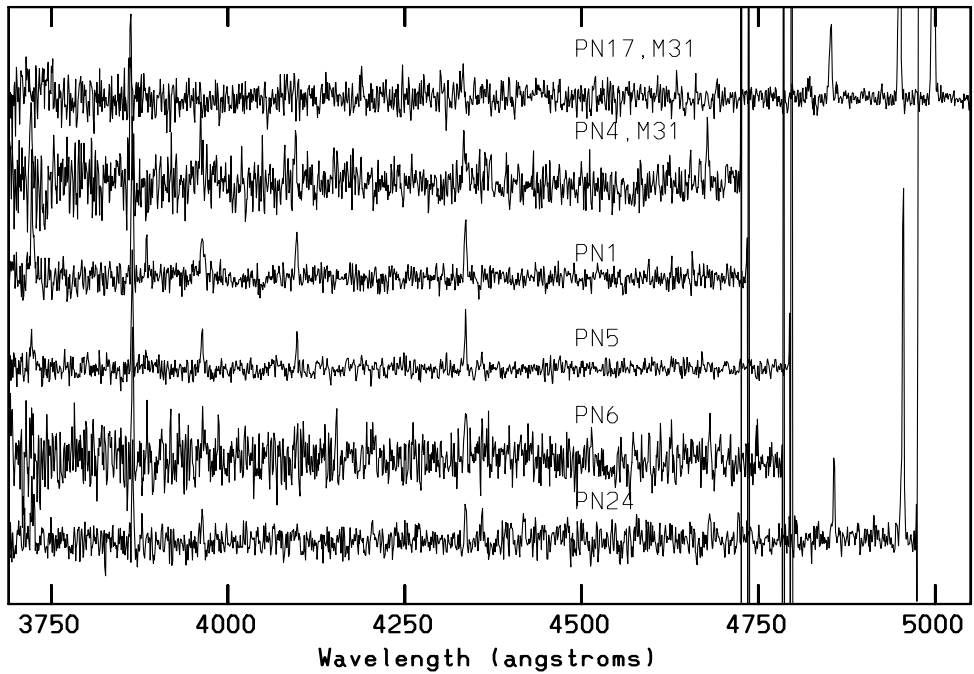,height=16.5cm,width=10.0cm,angle=90}
}
\caption{
The U900 spectra for PN24, PN6, PN5,  PN1,  PN4,
and  PN17  in  the \object{M32} field.  H$\beta$ defines the full  intensity
scale  for  PN24  and  PN17, but H$\gamma$ does  so  for  the  other
objects. The full useful wavelength range, 3690\AA\ to 5050\AA, is
displayed.  See Fig.~\ref{fig1b} for comments on individual objects.
}
\label{fig3b}
\end{figure*}

\begin{figure*}
%\resizebox{\hsize}{!}{\includegraphics{ap1to5.eps}}
\vspace{-0.5cm}
\hbox{
\epsfig{figure=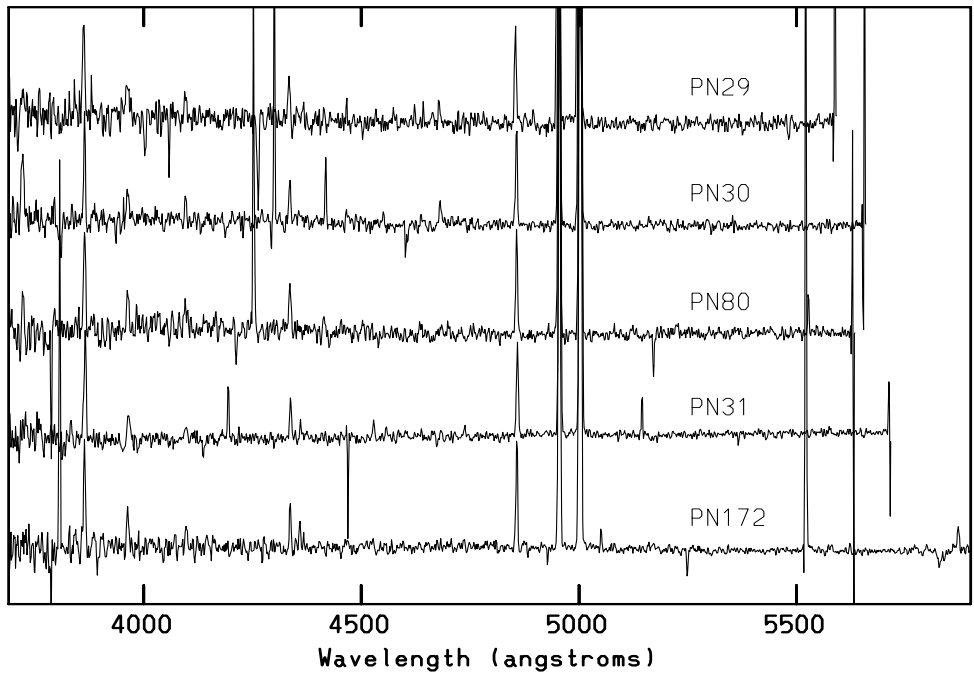,height=16.5cm,width=10.0cm,angle=90}
}
\caption{
The B600 spectra for PN172, PN31, PN80, PN30, and
PN29 in the \object{M31} bulge field.  The intensity scaling is set so
that  H$\beta$ occupies the full free intensity scale in all cases,
and the entire useful wavelength range is shown.
}
\label{fig4a}
\end{figure*}

\begin{figure*}
%\resizebox{\hsize}{!}{\includegraphics{ap6to10.eps}}
\vspace{-0.5cm}
\hbox{
\hspace{0cm}\epsfig{figure=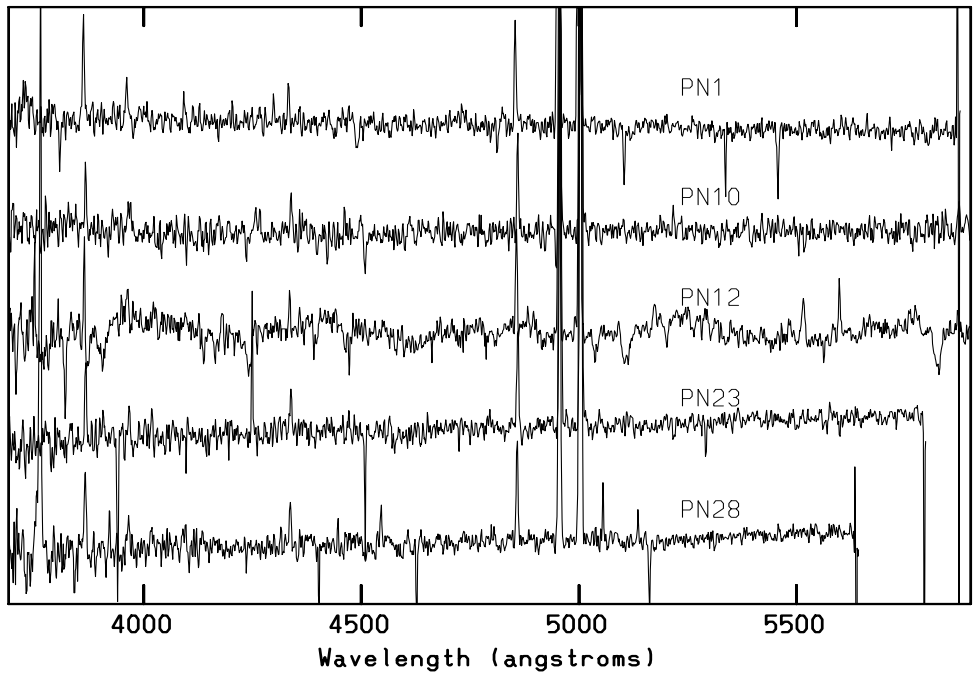,height=16.5cm,width=10.0cm,angle=90}
}
\caption{
The B600 spectra for PN28, PN23, PN12, PN10,  and
PN1  in  the  \object{M31} bulge field.  The intensity and  wavelength
scales  are  as  in  Fig. \ref{fig4a}.  Note  that  the  background
subtraction is poorer for PN12 than is normally the case.
}
\label{fig4b}
\end{figure*}

\begin{figure*}
%\resizebox{\hsize}{!}{\includegraphics{ap11to15.eps}}
\vspace{-0.5cm}
\hbox{
\hspace{0cm}\epsfig{figure=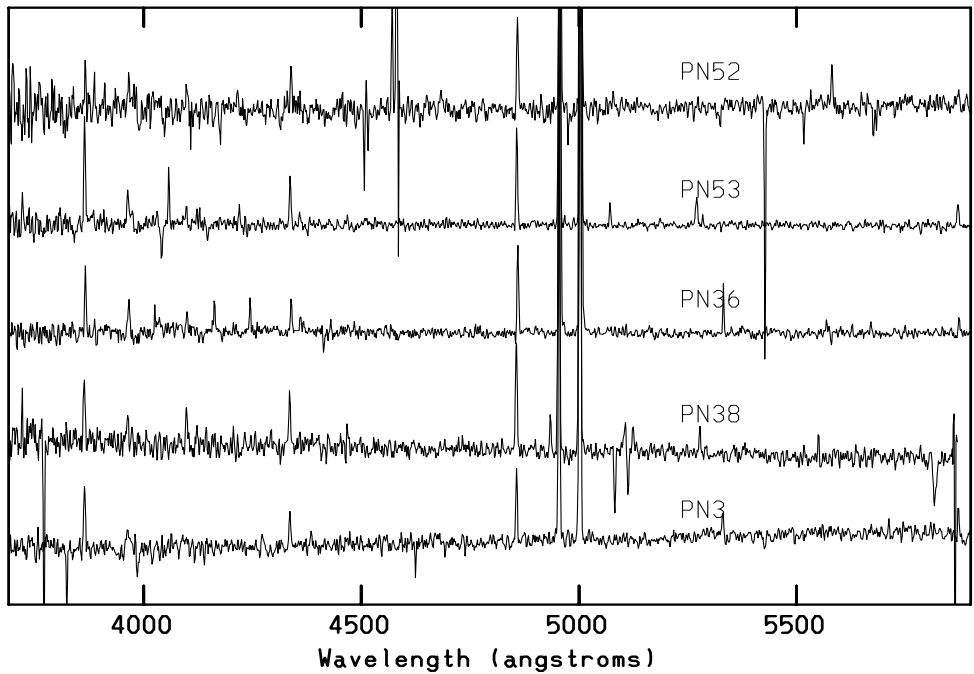,height=16.5cm,width=10.0cm,angle=90}
}
\caption{
The B600 spectra for PN3, PN38, PN36, PN53,  and
PN52  in  the \object{M31} bulge field.  The intensity and  wavelength
scales are as in Fig. \ref{fig4a}.
}
\label{fig4c}
\end{figure*}

\begin{figure*}
%\resizebox{\hsize}{!}{\includegraphics{ap16to20.eps}}
\vspace{-0.5cm}
\hbox{
\hspace{0cm}\epsfig{figure=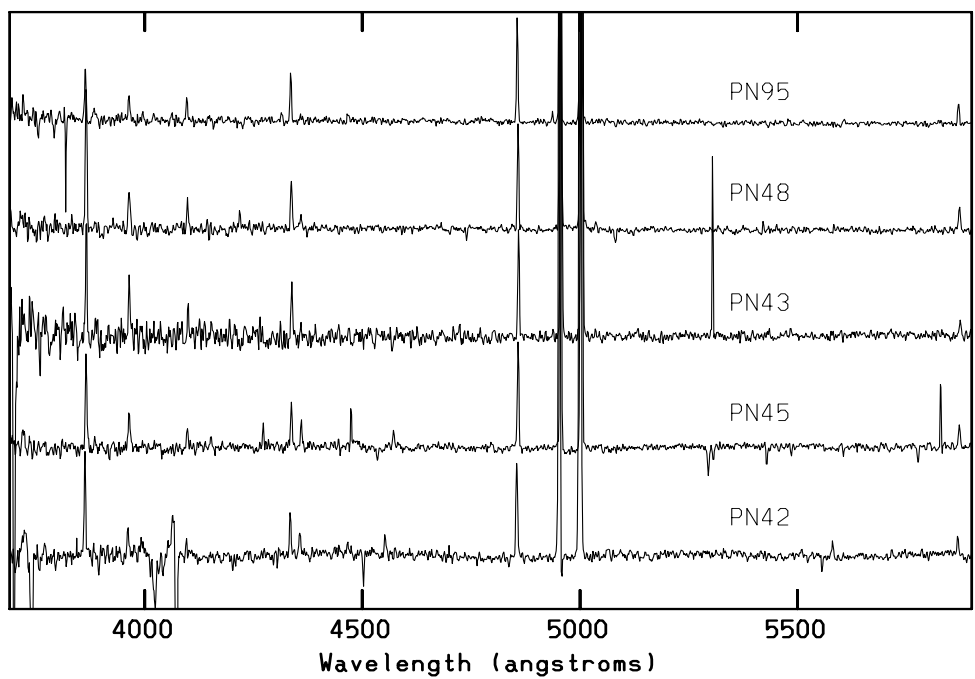,height=16.5cm,width=10.0cm,angle=90}
}
\caption{
The B600 spectra for PN42, PN45, PN43, PN48,  and
PN95  in  the \object{M31} bulge field.  The intensity and  wavelength
scales are as in Fig. \ref{fig4a}.
}
\label{fig4d}
\end{figure*}

\begin{figure*}
%\resizebox{\hsize}{!}{\includegraphics{ap21to25.eps}}
\vspace{-0.5cm}
\hbox{
\hspace{0cm}\epsfig{figure=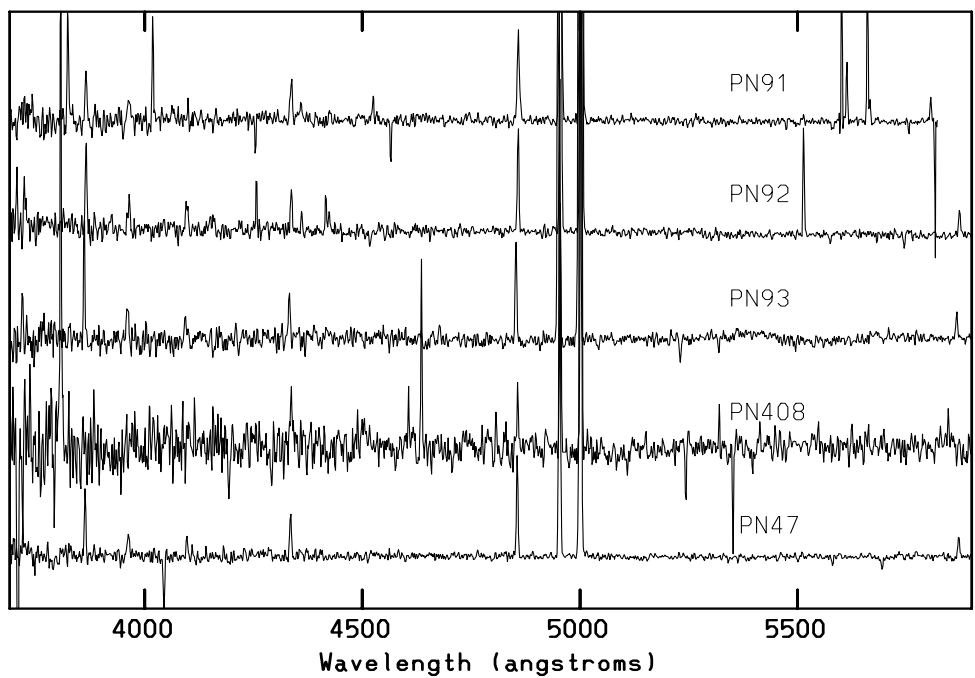,height=16.5cm,width=10.0cm,angle=90}
}
\caption{
The B600 spectra for PN47, PN408, PN93, PN92, and
PN91  in  the \object{M31} bulge field.  The intensity and  wavelength
scales are as in Fig. \ref{fig4a}.  Note that the signal-to-noise is
poor for the very faint object PN408.
}
\label{fig4e}
\end{figure*}

\begin{figure*}
%\resizebox{\hsize}{!}{\includegraphics{ap26to28.eps}}
\vspace{-0.5cm}
\hbox{
\hspace{0cm}\epsfig{figure=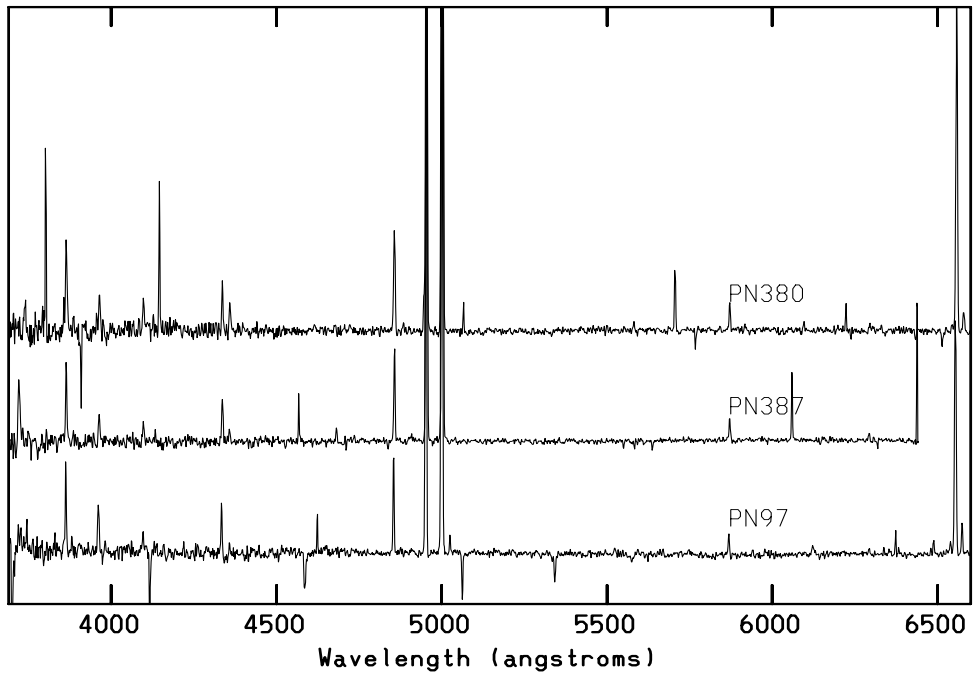,height=16.5cm,width=10.0cm,angle=90}
}
\caption{
The B600 spectra for PN97, PN387, and  PN380  in
the \object{M31} bulge field.  The intensity scaling is set so that H$\beta$
occupies the full free intensity scale in all cases, and  the
entire useful wavelength range is shown.  For PN97 and PN380,
though, the wavelength range extends to 6600\AA.
}
\label{fig4f}
\end{figure*}

Figures  \ref{fig1a} through \ref{fig3b} display  the  O300,
B600, and U900 spectra of the planetary nebulae in \object{M32}, while
Figures \ref{fig4a} through \ref{fig4f} display the
B600 spectra of the planetary  nebulae
in  the  bulge  of  \object{M31}.  The object designations  (Ciardullo
et al.
\cite{Ciardulloetal1989})
are shown next to the spectra.   Normally,  the
spectra  are scaled such that H$\beta$ occupies the full  intensity
scale,  so stronger lines from adjacent spectra overlap,  but
some of the U900 and B600 spectra are scaled such that H$\gamma$ and
H$\alpha$,  respectively,  occupy the full  intensity  scale.   This
scaling  allowed  the  best compromise in  demonstrating  the
signal-to-noise  for various lines and an assessment  of  the
background  sky and galaxy subtraction.  The full  wavelength
range  is  shown for the B600 and U900 spectra, but only  the
wavelength  range below 7350\AA\ is shown for the O300  spectra.
Cosmic rays were not removed unless they interfered with  the
measurement  of  line intensities, and  many  remain  in  the
spectra displayed in Figures \ref{fig1a} through \ref{fig4f}.

\section{Line intensities and reddening}

Tables \ref{table4} to \ref{table5} (at end)
list the adopted reddening-corrected line
intensity  ratios  and  reddening values  for  the  planetary
nebulae  in \object{M32} and the bulge of \object{M31}.   We  use
the  object  designations from Ciardullo et al.
(\cite{Ciardulloetal1989}).
The
line  intensities were measured using the software  described
by  McCall et al.
(\cite{McCalletal1985}).
The uncertainties  quoted
for the line ratios are $1\sigma$ uncertainties that incorporate the
uncertainties   in  both  the  line  and  H$\beta$   fluxes.    The
uncertainties  in the line fluxes include contributions  from
the  fit  to  the  line  itself and from  the  noise  in  the
continuum.   In  those  instances  where  there  is  no  line
intensity  value, but there is a line intensity  uncertainty,
e.g.,  \ion{He}{ii}\,$\lambda$4686  in PN5 in \object{M32}, the
\lq\lq uncertainty"  is  a  $2\sigma$
upper limit to the strength of undetected lines, and is based
upon the noise observed in the continuum.  Note that PN4  and
PN17  in the \object{M32} field have radial velocities indicating that
they  belong  to  the background disk of \object{M31} (Ford  \&  Jenner
\cite{FordJenner1975}).
The \ion{H}{ii} region in the background disk of \object{M31} that  we
observed  in the \object{M32} field is that denoted \ion{H}{ii} 1
by  Ford  \& Jenner
(\cite{FordJenner1975}).

For  the  planetary nebulae in \object{M32}, Tables \ref{table4}
to \ref{table4c} list  the reddening-corrected O300, B600, and
U900 line intensities, in addition  to  our  adopted  line
intensities.   The  adopted intensities   are  those  listed under
the   object   name. Generally, we adopted the U900 line
intensities in  the  blue and  the  O300 line intensities in the
red, with the dividing line being \ion{He}{ii}\,$\lambda$4686.
\ion{He}{ii}\,$\lambda$4686 is the only common exception to  this
rule.   For \ion{He}{ii}\,$\lambda$4686, we normally chose  the
line intensity  from the spectrum in which the line  was  measured
with the lowest relative error.

The reddening-corrected line intensities in Tables \ref{table4}
through \ref{table5} are related to those we observed via
$$\log\frac{I\left(\lambda\right)}{I\left({\mathrm
H}\beta\right)}= \log\frac{F\left(\lambda\right)}{F\left({\mathrm
H}\beta\right)} -0.4 E\left(B-V\right) \left(A\left(\lambda\right)
- A\left({\mathrm H}\beta\right)\right)$$ where  $F(\lambda)$ and
$I(\lambda)$ are the observed and reddening-corrected line
intensities, respectively, $E(B-V)$ is the reddening,  and
$A(\lambda)$ is the extinction for $E(B-V)=1.0$\,mag from the
reddening  law of Schild (\cite{Schild1977}). All  of the line
intensities  for the planetary  nebulae in \object{M32} in Tables
\ref{table4}, \ref{table4b}, and \ref{table4c} have been corrected
for reddening  using $E(B-V)$ determined from the O300
H$\alpha/\mathrm{H}\beta$  ratio. For  the U900 spectra that did
not extend to H$\beta$, we corrected intensities  relative  to
H$\gamma$ using the O300  reddening,  then adopted
$I(\mathrm{H}\gamma)/I(\mathrm{H}\beta)=0.47$. For the planetary
nebulae  in  the bulge  of  \object{M31},  we determined the
reddening from  the H$\alpha/\mathrm{H}\beta$ ratio  in  the two
cases when it was available, but used  the reddening calculated
from the $\mathrm{H}\gamma/\mathrm{H}\beta$ ratio otherwise.  In
all cases,  we  assumed intrinsic ratios of
$I(\mathrm{H}\alpha)/I(\mathrm{H}\beta)=2.85$ and
$I(\mathrm{H}\gamma)/I(\mathrm{H}\beta)=0.47$, which  are
appropriate  for  an  electron temperature  of  $10^4$\,K  and  an
electron  density  of $10^4\,{\rm cm}^{-3}$ (Osterbrock
\cite{Osterbrock1989}). The reddening uncertainties reflect  the
$1\sigma$ uncertainties in the H$\alpha$ or H$\gamma$ line
intensities.

Note that the line intensities for PN408 in \object{M31} are  not
corrected for reddening.  For this faint object, we  did  not
detect H$\gamma$, and H$\alpha$ fell outside our spectral window.

Since  our  reddenings  are based  upon  different  line
intensity ratios for different objects, we consider  them  in
greater  detail  before  proceeding.   All  of  our  H$\alpha$-based
reddenings  in Tables \ref{table4} through
\ref{table5} are positive.  The overwhelming
majority  of  our  H$\gamma$-based reddenings in Table \ref{table5} are also
either  positive or consistent with no reddening, but our  $1\sigma$
H$\gamma$  line intensity uncertainties do allow negative reddenings
in four cases (PN3, PN43, PN48, and PN53).  We considered not
using H$\gamma$ to determine the reddening, but rejected this option
for  four reasons.  First, for the four planetary nebulae  in
\object{M32}  for  which  we measured an H$\gamma$ intensity  from  the  B600
spectrum,  the  reddening-corrected  H$\gamma$  intensity  has   the
expected  value  of  approximately  47\%  that  of  H$\beta$   after
correcting  for  reddening using the O300 H$\alpha$  intensity.   In
these  four  cases,  then,  H$\alpha$ and  H$\gamma$  would  yield  similar
reddenings.   Second,  our  ultimate  aim  is  to   calculate
electron  temperatures and oxygen abundances from these  line
intensities.   If  we measured the intensity  of  [\ion{O}{iii}]$\lambda$4363
relative  to  H$\gamma$  and
[\ion{O}{iii}]$\lambda\lambda$4959,5007 relative  to  H$\beta$,  and
assumed  $I(\mathrm{H}\gamma)/I(\mathrm{H}\beta)=0.47$,
we would obtain final  intensities
for   the   [\ion{O}{iii}]   lines   that  would   be   statistically
indistinguishable  from  those  obtained  by  correcting  for
reddening  using  the  H$\gamma$  intensity.   Applying  a  negative
reddening  correction  does affect the  oxygen  abundance  we
derive  by reducing the [\ion{O}{ii}]$\lambda$3727 intensity, but this effect
has  less impact on the oxygen abundance than the uncertainty
in  the  electron temperature since there is so little oxygen
in  the  form of O${}^{+}$.  Third, forcing
$I(\mathrm{H}\gamma)/I(\mathrm{H}\beta)=0.47$  via  a
reddening  correction,  even if negative,  accounts  for  any
errors  in  the sensitivity calibration that might  otherwise
systematically  affect the [\ion{O}{iii}] lines  and  the  subsequent
oxygen  abundances.  Fourth, on average, our H$\alpha$- and H$\gamma$-based
reddenings  agree.   The  mean  H$\alpha$-based  reddening  for  all
objects  (both \object{M31} and \object{M32}) is
$E(B-V)=0.18\pm 0.04$\,mag, while  the
mean  H$\gamma$-based reddening for all of the planetary nebulae  in
the bulge of \object{M31} is $E(B-V)=0.18\pm 0.08$\,mag,
if negative reddening
values  are  included,  or  $E(B-V)=0.25\pm 0.06$\,mag,  if  negative
reddening values are set to zero (the uncertainties  are  the
standard errors in the means).  Thus, the reddenings computed
from  H$\alpha$  and H$\gamma$ are similar.  For comparison,
the foreground
reddening  to \object{M31} is $E(B-V)=0.093\pm 0.009$\,mag
(mean of McClure  \&
Racine
\cite{McClureRacine1969},
van den Bergh
\cite{vandenBergh1969},
and Burstein \& Heiles
\cite{BursteinHeiles1984}).
It  is  not  surprising  that  the  mean  reddening  for  the
planetary  nebulae  is 0.10\,mag greater  than  the  foreground
value, for planetary nebulae suffer additional reddening  due
to  internal dust and dust within \object{M31} and \object{M32}.
Consequently,
we have chosen to correct for \lq\lq reddening" even when $E(B-V)$ is
negative.

\section{Oxygen abundances}

\begin{table}
\caption[]{Oxygen Abundances in \object{M32}}
\label{table6}
\[
\begin{tabular}{llll}
\hline
\noalign{\smallskip}
Object & T$_e$            & 12+log(O/H)     & 12+log(O/H)     \\
       &                  & no ICF & KB94 ICF\,$^{\mathrm a}$ \\
       & (K)              & (dex)           & (dex, adopted)  \\
\noalign{\smallskip}
\hline
\noalign{\smallskip}
PN8    & $13000 \pm 1400$ & $8.36 \pm 0.14$ & $8.38 \pm 0.14$ \\
PN11   & $< 25300       $ & $> 7.25       $ & $> 7.31       $ \\
PN2    & $< 15700       $ & $> 8.17       $ & $> 8.27       $ \\
PN7    & $< 13400       $ & $> 8.08       $ & $> 8.10       $ \\
PN25   & $< 18600       $ & $> 8.02       $ & $> 8.08       $ \\
PN24   & $18200 \pm 2500$ & $8.07 \pm 0.13$ & $8.13 \pm 0.13$ \\
PN6    & $12700 \pm 2500$ & $8.46 \pm 0.27$ & $8.50 \pm 0.27$ \\
PN5    & $11800 \pm 1300$ & $8.43 \pm 0.16$ & $8.46 \pm 0.16$ \\
PN1    & $< 11500       $ & $> 8.39       $ & $> 8.41       $ \\
PN4${}^{\mathrm{b}}$ & $< 12200$ & $> 8.58$ & $> 8.72       $ \\
PN17${}^{\mathrm{b}}$& $< 18400$ & $> 7.84$ & $> 7.92       $ \\
\noalign{\smallskip}
\hline
\end{tabular}
\]
\begin{list}{}{}
\item[${}^{\mathrm{a}}$] Kingsburgh \& Barlow
\cite{KingsburghBarlow1994} ICF
\item[${}^{\mathrm{b}}$] PN4 and PN17 are background objects in the disk
of \object{M31} (Ford \& Jenner \cite{FordJenner1975}).
\end{list}
\end{table}

\begin{table}
\caption[]{Oxygen Abundances in \object{M31}} \label{table7}
\[
\begin{tabular}{llll}
\hline \noalign{\smallskip}
Object  & T$_e$            & 12+log(O/H)     & 12+log(O/H)    \\
        &                  & no ICF & KB94 ICF\,$^{\mathrm a}$\\
        & (K)              & (dex)           & (dex, adopted) \\
\noalign{\smallskip}
\hline
\noalign{\smallskip}
PN172   & $13100 \pm 1100$ & $8.44 \pm 0.10$ & $8.46 \pm 0.10$\\
PN31    & $10700 \pm 780 $ & $8.80 \pm 0.10$ & $8.80 \pm 0.10$\\
PN80    & $< 9500        $ & $> 8.91       $ & $> 8.91       $\\
PN30    & $< 9300        $ & $> 9.00       $ & $> 9.00       $\\
PN29    & $< 9800        $ & $> 8.91       $ & $> 8.91       $\\
PN28    & $< 8600        $ & $> 9.11       $ & $> 9.11       $\\
PN23    & $< 13200       $ & $> 8.30       $ & $> 8.30       $\\
PN12    & $13100 \pm 2000$ & $8.44 \pm 0.18$ & $8.51 \pm 0.18$\\
PN10    & $< 13600       $ & $> 8.22       $ & $> 8.24       $\\
PN1     & $< 10100       $ & $> 8.79       $ & $> 8.79       $\\
PN3     & $< 9400        $ & $> 8.96       $ & $> 8.96       $\\
PN38    & $< 10500       $ & $> 8.58       $ & $> 8.58       $\\
PN36    & $14600 \pm 1600$ & $8.12 \pm 0.12$ & $8.15 \pm 0.12$\\
PN53    & $9600 \pm 870  $ & $8.91 \pm 0.14$ & $8.92 \pm 0.14$\\
PN52    & $< 11300       $ & $> 8.53       $ & $> 8.64       $\\
PN42    & $12400 \pm 750 $ & $8.57 \pm 0.08$ & $8.59 \pm 0.08$\\
PN45    & $15100 \pm 940 $ & $8.16 \pm 0.07$ & $8.17 \pm 0.07$\\
PN43    & $< 10200       $ & $> 8.71       $ & $> 8.72       $\\
PN48    & $9500 \pm 640  $ & $8.95 \pm 0.11$ & $8.96 \pm 0.11$\\
PN95    & $< 9300        $ & $> 8.66       $ & $> 8.66       $\\
PN47    & $9600 \pm 1200 $ & $8.77 \pm 0.20$ & $8.78 \pm 0.20$\\
PN408   & $< 13000       $ & $> 8.52       $ & $> 8.52       $\\
PN93    & $< 10500       $ & $> 8.79       $ & $> 8.80       $\\
PN92    & $11800 \pm 1100$ & $8.56 \pm 0.12$ & $8.57 \pm 0.12$\\
PN97    & $9400 \pm 1000 $ & $8.87 \pm 0.18$ & $8.88 \pm 0.18$\\
PN91    & $12700 \pm 1400$ & $8.38 \pm 0.13$ & $8.38 \pm 0.13$\\
PN387   & $10600 \pm 810 $ & $8.71 \pm 0.12$ & $8.74 \pm 0.12$\\
PN380   & $13900 \pm 1100$ & $8.38 \pm 0.09$ & $8.39 \pm 0.09$\\
\noalign{\smallskip}
\hline
\end{tabular}
\]
\begin{list}{}{}
\item[${}^{\mathrm{a}}$] Kingsburgh \& Barlow
\cite{KingsburghBarlow1994} ICF
\end{list}
\end{table}

Tables \ref{table6} and \ref{table7} present the electron
temperatures and the oxygen abundances for the planetary nebulae
in \object{M32} and in the bulge  of  \object{M31}, respectively.
We only observed two ionization stages of oxygen, O${}^{+}$  and
O${}^{++}$.  We accounted for unseen stages in our oxygen
abundance calculations using the ionization correction factors
(ICF) computed according to the prescription of Kingsburgh \&
Barlow (\cite{KingsburghBarlow1994}), which employs the line
intensities of \ion{He}{ii}\,$\lambda$4686 and
\ion{He}{i}\,$\lambda$5876  to correct  for unseen  ionization
stages of oxygen.  Further details may be found in Stasi\'nska et
al. (\cite{Stasinskaetal1998}).  Tables \ref{table6} and
\ref{table7} present two oxygen abundance calculations.  The
abundances in column 3 are simply the sum of the O${}^{+}$ and
O${}^{++}$ ionic abundances. The abundances in column 4 are those
from column 3 corrected  for the ICF.   The ICF is normally small
because \ion{He}{ii}\,$\lambda$4686  is weak. The oxygen
abundances in column 4 will be adopted  in future work.

In  calculating  the oxygen abundances,  we  assumed  an
electron  density  of $4000\,{\rm cm}^{-3}$ in all cases.   With
electron
densities of $1\,{\rm cm}^{-3}$ and $2\,10^{4}{\rm cm}^{-3}$,
the oxygen abundance changes
by  a maximum of $-0.02$\,dex and $+0.07$\,dex, respectively, for the
planetary  nebulae in \object{M31}, and by a maximum of  $-0.03$\,dex  and
$+0.11$\,dex, respectively, for the planetary nebulae in \object{M32}.

In instances where only upper limits to intensities were
available,  we adopted the following approach.  When  we  had
upper  limits for the intensities of the helium  lines  these
limits were used to calculate the ICF.  If we did not observe
He~{\sc i}\,$\lambda$5876  (because  it was outside our spectral
window),  we
made no correction for unseen stages of oxygen regardless  of
the  intensity  of He~{\sc ii}\,$\lambda$4686.  (Only in two cases,
PN29  and
PN30  in  \object{M31},  did  we detect \ion{He}{ii}\,$\lambda$4686 when
\ion{He}{i}\,$\lambda$5876  was
outside  our  spectral window.)  When we only  had  an  upper
limit  to [\ion{O}{iii}]$\lambda$4363, we used this to derive an
upper limit
to  the electron temperature, and this temperature limit  was
then  used  to derive a lower limit to the oxygen  abundance.
In  these  instances, we did not compute an error for  either
the  electron temperature or the oxygen abundance,  and  have
indicated  the results listed in Tables \ref{table6} and
\ref{table7} as  limits.
When  we  had an upper limit for [\ion{O}{ii}]$\lambda$3727, we
adopted  this
limiting intensity for the line.  In this case, the O${}^{+}$  ionic
abundance  is  over-estimated, but its  contribution  to  the
total oxygen abundance was normally small.

Our  uncertainties  for  the electron  temperatures  and
oxygen  abundances reflect the uncertainties  in  the  [\ion{O}{iii}]
line   intensities   alone.   As  noted  earlier,   reddening
introduces  a  further uncertainty through  its  effect  upon
[\ion{O}{ii}]$\lambda$3727,  but  this  has less influence
upon  the  oxygen
abundance  than the uncertainty in the electron  temperature.
The  electron temperature uncertainty that we quote is simply
the  temperature range permitted by the ($1\sigma$) limiting  values
of  the  [\ion{O}{iii}]  line  intensities.   Similarly,  our  oxygen
abundance  uncertainties  are  derived  from  the  abundances
calculated   using  the  extreme  values  of   the   electron
temperature.

\section{Discussion}

In Tables \ref{table6} and \ref{table7}, we derive oxygen
abundances  for
approximately half of the planetary nebulae we observed.  For
the  rest,  we  derive  lower limits.   Many  of  the  oxygen
abundance  limits,  however, are very  useful.   Six  of  the
fourteen  temperature limits in Table \ref{table7} are below
$10^4$\,K,  and
one  is  even  below  9000\,K.  If we  separate  the  planetary
nebulae  in  Table  \ref{table7}  on  the basis  of  whether
they  have
temperatures   or   temperature  limits,  the   mean   oxygen
abundances  of  the  two sets differ at  the  92\% confidence
level, with the set of objects with temperature limits having
a  higher mean oxygen abundance by at least 0.11\,dex.  Table
\ref{table7}
shows  clearly that we are able to measure oxygen  abundances
up  to  approximately  the solar value
($12+\log(\mathrm{O}/\mathrm{H})=8.93$\,dex; Anders  \&  Grevesse
\cite{AndersGrevesse1989}).
Since the Loral3  CCD  has  only
modest  sensitivity  at [\ion{O}{iii}]$\lambda$4363, the  quantum  efficiency
being  about 22\%, these results are by no means the limit  of
what is possible with 4m-class telescopes.

In several companion papers,  we shall exploit the
spectroscopic observations of planetary nebulae in \object{M32}
and in the bulge of \object{M31} in several ways.  First, we
intend to study the evolution of  the planetary nebulae in these
galaxies relative to those in  the Milky Way and the
\object{Magellanic Clouds}.  We also intend to investigate the
chemical evolution of \object{M32} and the  bulges of \object{M31}
and the \object{Milky Way} individually as well as the chemical
evolution of DHGs as a class.

\begin{acknowledgements}

We  would like to thank George Jacoby for making several very
helpful  comments concerning a preliminary  version  of this
paper, and for making further helpful suggestions as referee.  MGR
would like to thank the Natural Sciences and Engineering  Research
Council of Canada and Marshall McCall for their  financial support
while this research was being done. MGR also thanks the Physics
and Astronomy Department at York University for its hospitality
and computing facilities while this work was being finished.  MLM
thanks the Natural Sciences and Engineering Research Council of
Canada  for its continuing support.

\end{acknowledgements}

%\end{document}

\begin{table*}
\caption[]{Line Intensities for PNe in \object{M32}}
\label{table4}
\[
\begin{array}{l*{8}r}
\hline
\noalign{\smallskip}
{\rm Wavelength} &     {\rm PN5}       &      {\rm O300}       &     {\rm B600}        &    %%@
{\rm U900}         & {\rm PN1}             & {\rm O300}            & {\rm B600}            & %%@
{\rm U900}             \\
\noalign{\smallskip}
3727^{\mathrm a}     & 29.0    \pm 8.3   &                 &                 & 29.0    \pm %%@
8.3   & 83      \pm 13    &                 &                 & 83      \pm 13     \\
3868.76  & 94.0    \pm 9.9   &                 & 100     \pm 19    & 94.0    \pm 9.9   & 110     %%@
\pm 12    &                 & 78      \pm 13    & 110     \pm 12     \\
3967.47  & 48.6    \pm 9.3   &                 &                 & 48.6    \pm 9.3   & 49      %%@
\pm 11    &                 &                 & 49      \pm 11     \\
4101.737 & 29.8    \pm 5.8   &                 &                 & 29.8    \pm 5.8   & 33.3    %%@
\pm 5.8   &                 & 27.1    \pm 8.5   & 33.3    \pm 5.8    \\
4340.468 & 47.0    \pm 6.6   &                 &         \pm 13    & 47.0    \pm 6.6   & %%@
47.0    \pm 6.8   & 48.6    \pm 8.8   & 42.2    \pm 5.9   & 47.0    \pm 6.8    \\
4363.21  & 14.0    \pm 4.4   &                 &         \pm 13    & 14.0    \pm 4.4   &         %%@
\pm 10    &         \pm 16    &         \pm 9     &         \pm 10     \\
4685.682 &         \pm 17    &         \pm 17    &         \pm 14    & 10.0    \pm 3.3   & %%@
10.8    \pm 3.7   &         \pm 7     & 10.8    \pm 3.7   &         \pm 11     \\
4861.332 & 100.0   \pm 5.4   & 100.0   \pm 5.4   & 100.0   \pm 8.8   &                 & %%@
100.0   \pm 4.0   & 100.0   \pm 4.0   & 100.0   \pm 4.8   &                  \\
4958.92  & 431.7   \pm 5.0   & 431.7   \pm 5.0   & 448     \pm 26    &                 & %%@
337.8   \pm 4.6   & 337.8   \pm 4.6   & 349     \pm 11    &                  \\
5006.85  & 1291.8  \pm 6.9   & 1291.8  \pm 6.9   & 1343    \pm 71    &                 & %%@
994.1   \pm 6.7   & 994.1   \pm 6.7   & 1052    \pm 29    &                  \\
5875.666 & 16.0    \pm 4.0   & 16.0    \pm 4.0   & 18.8    \pm 4.9   &                 & %%@
21.2    \pm 1.6   & 21.2    \pm 1.6   & 22.3    \pm 2.6   &                  \\
6300.32  &         \pm 6     &         \pm 6     &                 &                 &         %%@
\pm 2     &         \pm 2     &                 &                  \\
6312.06  & 10.4    \pm 3.6   & 10.4    \pm 3.6   &                 &                 &         %%@
\pm 2     &         \pm 2     &                 &                  \\
6363.81  &         \pm 6     &         \pm 6     &                 &                 &         %%@
\pm 2     &         \pm 2     &                 &                  \\
6548.06  & 29.6    \pm 3.2   & 29.6    \pm 3.2   &                 &                 & 76.7    %%@
\pm 2.1   & 76.7    \pm 2.1   &                 &                  \\
6562.817 & 285.0   \pm 4.0   & 285.0   \pm 4.0   &                 &                 & 285.0   %%@
\pm 2.7   & 285.0   \pm 2.7   &                 &                  \\
6583.39  & 102.7   \pm 3.5   & 102.7   \pm 3.5   &                 &                 & 230.6   %%@
\pm 2.5   & 230.6   \pm 2.5   &                 &                  \\
6716.42  &                 &                 &                 &                 & 10.21   %%@
\pm 0.92  & 10.21   \pm 0.92  &                 &                  \\
6730.72  &                 &                 &                 &                 & 17.08   %%@
\pm 0.99  & 17.08   \pm 0.99  &                 &                  \\
7065.277 & 13.0    \pm 4.1   & 13.0    \pm 4.1   &                 &                 & 3.5     %%@
\pm 1.0   & 3.5     \pm 1.0   &                 &                  \\
7135.8   & 30.2    \pm 4.8   & 30.2    \pm 4.8   &                 &                 & 18.0    %%@
\pm 1.3   & 18.0    \pm 1.3   &                 &                  \\
7325     & 10.4    \pm 3.3   & 10.4    \pm 3.3   &                 &                 & 3.11    %%@
\pm 0.56  & 3.11    \pm 0.56  &                 &                  \\
9069     & 12.7    \pm 3.1   & 12.7    \pm 3.1   &                 &                 &                 %%@
&                 &                 &                  \\
E(B-V)   & 0.01    \pm 0.01  &                 &                 &                 & 0.11    %%@
\pm 0.01  &                 &                 &                  \\
\noalign{\smallskip}
{\rm Wavelength} & {\rm PN4, M31}      & {\rm O300}            & {\rm B600}            & %%@
{\rm U900}            & {\rm PN17, M31}       & {\rm O300}            & {\rm B600}            %%@
& {\rm U900}             \\
\noalign{\smallskip}
3727^{\mathrm a}     & 106     \pm 36    &                 &                 & 106     \pm %%@
36    &         \pm 64    &                 &                 &         \pm 64     \\
3868.76  & 169     \pm 32    &                 &                 & 169     \pm 32    & 79      %%@
\pm 22    &                 &                 & 79      \pm 22     \\
3967.47  & 53      \pm 15    &                 &                 & 53      \pm 15    &                 %%@
&                 &                 &                  \\
4101.737 & 50      \pm 17    &                 &                 & 50      \pm 17    &         %%@
\pm 34    &                 &                 &         \pm 34     \\
4340.468 & 47      \pm 13    &                 &                 & 47      \pm 13    & 41      %%@
\pm 13    &                 &                 & 41      \pm 13     \\
4363.21  &         \pm 22    &                 &                 &         \pm 22    &         %%@
\pm 29    &                 &                 &         \pm 29     \\
4685.682 & 51      \pm 11    & 51      \pm 11    & 37      \pm 10    & 55      \pm 14    &         %%@
\pm 22    &         \pm 26    &         \pm 25    &         \pm 22     \\
4861.332 & 100     \pm 11    & 100     \pm 11    & 100     \pm 12    &                 & 100     %%@
\pm 12    & 100     \pm 12    & 100     \pm 14    & 100     \pm 10     \\
4958.92  & 637     \pm 10    & 637     \pm 10    & 703     \pm 50    &                 & 337     %%@
\pm 12    & 337     \pm 12    & 384     \pm 38    & 253     \pm 18     \\
5006.85  & 1916    \pm 14    & 1916    \pm 14    & 2137    \pm 149   &                 & 993     %%@
\pm 16    & 993     \pm 16    & 1176    \pm 110   & 807     \pm 48     \\
5875.666 & 10.9    \pm 4.7   &         \pm 4     & 10.9    \pm 4.7   &                 & %%@
13.0    \pm 3.4   & 13.0    \pm 3.4   & 19.3    \pm 4.9   &                  \\
6300.32  &         \pm 6     &         \pm 6     &                 &                 &         %%@
\pm 4     &         \pm 4     &                 &                  \\
6312.06  &         \pm 6     &         \pm 6     &                 &                 &         %%@
\pm 4     &         \pm 4     &                 &                  \\
6363.81  &         \pm 6     &         \pm 6     &                 &                 &         %%@
\pm 4     &         \pm 4     &                 &                  \\
6548.06  & 15.6    \pm 4.0   & 15.6    \pm 4.0   &                 &                 & 26.3    %%@
\pm 2.4   & 26.3    \pm 2.4   & 29.6    \pm 5.3   &                  \\
6562.817 & 285.0   \pm 5.7   & 285.0   \pm 5.7   &                 &                 & 285.0   %%@
\pm 4.1   & 285.0   \pm 4.1   & 349     \pm 33    &                  \\
6583.39  & 45.4    \pm 4.5   & 45.4    \pm 4.5   &                 &                 & 77.3    %%@
\pm 2.7   & 77.3    \pm 2.7   & 91.7    \pm 9.8   &                  \\
7135.8   & 21.8    \pm 3.1   & 21.8    \pm 3.1   &                 &                 & 22.5    %%@
\pm 4.1   &                 & 22.5    \pm 4.1   &                  \\
7325     & 14.4    \pm 5.4   & 14.4    \pm 5.4   &                 &                 & 14.2    %%@
\pm 1.9   & 14.2    \pm 1.9   &                 &                  \\
E(B-V)   & 0.15    \pm 0.02  &                 &                 &                 & 0.57    %%@
\pm 0.01  &                 &                 &                  \\
\noalign{\smallskip}
\hline
\end{array}
\]
\begin{list}{}{}
\item[$^{\mathrm a}$] \lq\lq 3727" denotes the sum of
[\ion{O}{ii}]$\lambda\lambda$3726,3729
\end{list}
\end{table*}

\begin{table*}
\caption[]{Line Intensities for PNe in \object{M32} (continued)}
\label{table4b}
\[
\begin{array}{l*{8}r}
\hline
\noalign{\smallskip}
{\rm Wavelength} & {\rm PN8}           & {\rm O300}            & {\rm B600}            & %%@
{\rm U900}            & {\rm PN2}             & {\rm O300}            & {\rm B600}            %%@
& {\rm U900}             \\
\noalign{\smallskip}
3726.05  & 38.1    \pm 9.0   &                 &                 & 38.1    \pm 9.0   &         %%@
\pm 35    &                 &                 &         \pm 35     \\
3728.8   & 41.9    \pm 9.1   &                 &                 & 41.9    \pm 9.1   & 85      %%@
\pm 22    &                 &                 & 85      \pm 22     \\
3868.76  & 121     \pm 10    &                 &                 & 121     \pm 10    & 98      %%@
\pm 17    &                 &                 & 98      \pm 17     \\
3967.47  & 62      \pm 13    &                 &                 & 62      \pm 13    & 60      %%@
\pm 17    &                 &                 & 60      \pm 17     \\
4101.737 & 24.1    \pm 7.6   &                 &                 & 24.1    \pm 7.6   &         %%@
\pm 21    &                 &                 &         \pm 21     \\
4340.468 & 59.7    \pm 6.8   &                 & 42.4    \pm 6.0   & 59.7    \pm 6.8   & 73      %%@
\pm 17    &                 &                 & 73      \pm 17     \\
4363.21  & 19.1    \pm 5.3   &                 & 13.4    \pm 4.7   & 19.1    \pm 5.3   &         %%@
\pm 31    &                 &                 &         \pm 31     \\
4685.682 &         \pm 11    & 5.3     \pm 2.5   &         \pm 8     &         \pm 11    & %%@
22.8    \pm 7.7   & 16.5    \pm 7.9   & 44      \pm 11    & 22.8    \pm 7.7    \\
4861.332 & 100.0   \pm 4.6   & 100.0   \pm 4.6   & 100.0   \pm 6.0   & 100.0   \pm 7.1   & %%@
100     \pm 10    & 100     \pm 10    & 100     \pm 13    & 100     \pm 11     \\
4958.92  & 462.5   \pm 7.6   & 462.5   \pm 7.6   & 458     \pm 17    & 434     \pm 18    & %%@
508.7   \pm 7.3   & 508.7   \pm 7.3   & 723     \pm 61    & 520     \pm 36     \\
5006.85  & 1386    \pm 11    & 1386    \pm 11    & 1364    \pm 48    & 1334    \pm 52    & %%@
1461.1  \pm 9.5   & 1461.1  \pm 9.5   & 2109    \pm 172   & 1602    \pm 100    \\
5875.666 & 16.6    \pm 1.1   & 16.6    \pm 1.1   & 18.2    \pm 2.2   &                 & 8.3     %%@
\pm 2.5   & 8.3     \pm 2.5   & 16.4    \pm 5.6   &                  \\
6300.32  & 5.8     \pm 1.4   & 5.8     \pm 1.4   &                 &                 & 8.6     %%@
\pm 2.6   & 8.6     \pm 2.6   &                 &                  \\
6312.06  & 3.2     \pm 1.3   & 3.2     \pm 1.3   &                 &                 &         %%@
\pm 3     &         \pm 3     &                 &                  \\
6363.81  &         \pm 2     &         \pm 2     &                 &                 &         %%@
\pm 3     &         \pm 3     &                 &                  \\
6548.06  & 57.0    \pm 1.2   & 57.0    \pm 1.2   & 51.0    \pm 3.7   &                 & %%@
30.9    \pm 2.0   & 30.9    \pm 2.0   & 50.6    \pm 8.3   &                  \\
6562.817 & 285.0   \pm 1.9   & 285.0   \pm 1.9   & 275.4   \pm 10    &                 & %%@
285.0   \pm 2.7   & 285.0   \pm 2.7   & 414     \pm 35    &                  \\
6583.39  & 168.2   \pm 1.5   & 168.2   \pm 1.5   & 158.2   \pm 6.7   &                 & %%@
102.4   \pm 2.3   & 102.4   \pm 2.3   & 159     \pm 15    &                  \\
6716.42  & 11.2    \pm 1.1   & 11.2    \pm 1.1   & 9.8     \pm 1.9   &                 & 9.0     %%@
\pm 1.8   & 9.0     \pm 1.8   &         \pm 13    &                  \\
6730.72  & 17.1    \pm 1.2   & 17.1    \pm 1.2   &                 &                 & 14.0    %%@
\pm 1.9   & 14.0    \pm 1.9   &                 &                  \\
7065.277 & 5.75    \pm 0.71  & 5.75    \pm 0.71  &                 &                 &                 %%@
&                 &                 &                  \\
7135.8   & 17.66   \pm 0.84  & 17.66   \pm 0.84  & 15.3    \pm 2.7   &                 & %%@
19.7    \pm 2.0   & 19.7    \pm 2.0   & 21.2    \pm 6.0   &                  \\
7325     & 2.36    \pm 0.57  & 2.36    \pm 0.57  &                 &                 &                 %%@
&                 &                 &                  \\
9069     & 13.2    \pm 1.2   & 13.2    \pm 1.2   &                 &                 & 20.1    %%@
\pm 3.3   & 20.1    \pm 3.3   &                 &                  \\
9531.8   & 34.0    \pm 3.7   & 34.0    \pm 3.7   &                 &                 & 32.2    %%@
\pm 4.7   & 32.2    \pm 4.7   &                 &                  \\
E(B-V)   & 0.20    \pm 0.01  &                 &                 &                 & 0.06    %%@
\pm 0.01  &                 &                 &                  \\
\noalign{\smallskip}
{\rm Wavelength} &     {\rm PN11}      &       {\rm O300}      &       {\rm B600}      &       %%@
{\rm U900}      & {\rm PN7}             & {\rm O300}            & {\rm B600}            & %%@
{\rm U900}             \\
\noalign{\smallskip}
3726.05  & 123     \pm 20    &                 &                 & 123     \pm 20    & 25.2    %%@
\pm 8.2   &                 &                 & 25.2    \pm 8.2    \\
3728.8   & 119     \pm 20    &                 &                 & 119     \pm 20    &         %%@
\pm 11    &                 &                 &         \pm 11     \\
3868.76  & 21.0    \pm 8.3   &                 &                 & 21.0    \pm 8.3   & 48      %%@
\pm 10    &                 & 49      \pm 17    & 48      \pm 10     \\
3967.47  & 16.5    \pm 6.8   &                 &                 & 16.5    \pm 6.8   & 21.3    %%@
\pm 5.4   &                 &                 & 21.3    \pm 5.4    \\
4101.737 & 16.9    \pm 8.3   &                 &                 & 16.9    \pm 8.3   & 24.8    %%@
\pm 8.7   &                 &                 & 24.8    \pm 8.7    \\
4340.468 & 21.1    \pm 9.4   &                 &         \pm 47    & 21.1    \pm 9.4   & %%@
47.0    \pm 8.4   &                 & 38.5    \pm 6.8   & 47.0    \pm 8.4    \\
4363.21  &         \pm 14    &                 &                 &         \pm 14    &         %%@
\pm 12    &                 &         \pm 9     &         \pm 12     \\
4685.682 &         \pm 16    &         \pm 15    &         \pm 21    &         \pm 16    & %%@
11.9    \pm 4.8   & 11.9    \pm 4.8   &         \pm 9     &                  \\
4861.332 & 100.0   \pm 9.2   & 100.0   \pm 9.2   & 100     \pm 16    & 100.0   \pm 9.5   & %%@
100.0   \pm 5.5   & 100.0   \pm 5.5   & 100.0   \pm 5.9   &                  \\
4958.92  & 91.4    \pm 6.5   & 91.4    \pm 6.5   & 106     \pm 14    & 94.6    \pm 9.2   & %%@
274.5   \pm 4.3   & 274.5   \pm 4.3   & 282     \pm 11    &                  \\
5006.85  & 298.4   \pm 8.7   & 298.4   \pm 8.7   & 320     \pm 34    & 263     \pm 17    & %%@
805.8   \pm 5.9   & 805.8   \pm 5.9   & 863     \pm 31    &                  \\
5875.666 & 17.1    \pm 3.0   & 17.1    \pm 3.0   & 15.2    \pm 6.9   &                 & %%@
14.0    \pm 1.4   & 14.0    \pm 1.4   & 12.5    \pm 2.0   &                  \\
6300.32  &         \pm 4     &         \pm 4     &                 &                 &                 %%@
&                 &                 &                  \\
6312.06  &         \pm 4     &         \pm 4     &                 &                 &                 %%@
&                 &                 &                  \\
6363.81  &         \pm 4     &         \pm 4     &                 &                 &                 %%@
&                 &                 &                  \\
6548.06  & 60.3    \pm 2.5   & 60.3    \pm 2.5   & 75.5    \pm 12.2  &                 & 5.7     %%@
\pm 1.2   & 5.7     \pm 1.2   &                 &                  \\
6562.817 & 285.0   \pm 5.1   & 285.0   \pm 5.1   & 315     \pm 33    &                 & %%@
285.0   \pm 2.1   & 285.0   \pm 2.1   &                 &                  \\
6583.39  & 178.7   \pm 3.7   & 178.7   \pm 3.7   & 199     \pm 23    &                 & %%@
12.3    \pm 1.3   & 12.3    \pm 1.3   &                 &                  \\
6716.42  & 12.0    \pm 2.0   & 12.0    \pm 2.0   &         \pm 13    &                 &         %%@
\pm 3     &         \pm 3     &                 &                  \\
6730.72  & 16.3    \pm 2.1   & 16.3    \pm 2.1   &                 &                 &         %%@
\pm 3     &         \pm 3     &                 &                  \\
7065.277 &                 &                 &                 &                 & 6.4     %%@
\pm 1.1   & 6.4     \pm 1.1   &                 &                  \\
7135.8   & 8.5     \pm 2.2   & 8.5     \pm 2.2   &                 &                 & 5.5     %%@
\pm 1.0   & 5.5     \pm 1.0   &                 &                  \\
7325     & 5.7     \pm 2.0   & 5.7     \pm 2.0   &                 &                 &                 %%@
&                 &                 &                  \\
9069     & 9.3     \pm 3.1   & 9.3     \pm 3.1   &                 &                 &                 %%@
&                 &                 &                  \\
E(B-V)   & 0.05    \pm 0.02  &                 &                 &                 & 0.03    %%@
\pm 0.01  &                 &                 &                  \\
\noalign{\smallskip}
\hline
\end{array}
\]
\end{table*}

\begin{table*}
\caption[]{Line Intensities for PNe in \object{M32} (continued)}
\label{table4c}
\[
\begin{array}{l*{8}r}
\hline
\noalign{\smallskip}
{\rm Wavelength} & {\rm H~II~\#1} & {\rm O300}            & {\rm B600}            & {\rm %%@
U900}            & {\rm PN24}            & {\rm O300}            & {\rm B600}            & %%@
{\rm U900}             \\
\noalign{\smallskip}
3726.05  & 98      \pm 33    &                 &                 & 98      \pm 33    & 73      %%@
\pm 15    &                 &                 & 73      \pm 15     \\
3727^{\mathrm a}     &                 &                 & 148     \pm 40    &                 %%@
&                 &                 &                 &                  \\
3728.8   & 58      \pm 27    &                 &                 & 58      \pm 27    & 48      %%@
\pm 13    &                 &                 & 48      \pm 13     \\
3868.76  & 26      \pm 14    &                 &                 & 26      \pm 14    & 134     %%@
\pm 20    &                 & 135     \pm 36    & 134     \pm 20     \\
3967.47  &                 &                 &                 &                 & 54      %%@
\pm 15    &                 &                 & 54      \pm 15     \\
4101.737 &         \pm 21    &                 &                 &         \pm 21    &         %%@
\pm 30    &                 &                 &         \pm 30     \\
4340.468 & 47      \pm 16    &                 &         \pm 13    & 47      \pm 16    & 67      %%@
\pm 13    &                 & 46      \pm 16    & 67      \pm 13     \\
4363.21  &         \pm 23    &                 &         \pm 13    &         \pm 23    & 47      %%@
\pm 12    &                 &                 & 47      \pm 12     \\
4685.682 &         \pm 17    &         \pm 8     &         \pm 17    &                 & 39      %%@
\pm 12    &         \pm 34    & 32      \pm 13    & 39      \pm 12     \\
4861.332 & 100.0   \pm 6.5   & 100.0   \pm 6.5   & 100     \pm 11    &                 & 100     %%@
\pm 19    & 100     \pm 19    & 100     \pm 21    & 100     \pm 14     \\
4958.92  & 63.0    \pm 5.8   & 63.0    \pm 5.8   & 73.1    \pm 7.5   &                 & 572     %%@
\pm 23    & 572     \pm 23    & 541     \pm 71    & 465     \pm 41     \\
5006.85  & 205.5   \pm 7.5   & 205.5   \pm 7.5   & 215     \pm 15    &                 & %%@
1654    \pm 30    & 1654    \pm 30    & 1585    \pm 202   &                  \\
5875.666 & 11.5    \pm 1.8   & 11.5    \pm 1.8   &                 &                 & 22      %%@
\pm 10    & 64      \pm 14    & 22      \pm 10    &                  \\
6300.32  &         \pm 4     &         \pm 4     &                 &                 &         %%@
\pm 27    &         \pm 27    &                 &                  \\
6312.06  &         \pm 4     &         \pm 4     &                 &                 &         %%@
\pm 27    &         \pm 27    &                 &                  \\
6363.81  &         \pm 4     &         \pm 4     &                 &                 &         %%@
\pm 27    &         \pm 27    &                 &                  \\
6548.06  & 28.3    \pm 2.3   & 28.3    \pm 2.3   &                 &                 & 94.7    %%@
\pm 9.7   & 94.7    \pm 9.7   & 63.0    \pm 15.1  &                  \\
6562.817 & 285.0   \pm 5.5   & 285.0   \pm 5.5   &                 &                 & 285     %%@
\pm 11    & 285     \pm 11    & 241     \pm 34    &                  \\
6583.39  & 23.6    \pm 2.0   & 23.6    \pm 2.0   &                 &                 & 268     %%@
\pm 11    & 268     \pm 11    & 214     \pm 31    &                  \\
6716.42  &                 &                 &                 &                 & 9.5     %%@
\pm 2.5   & 9.5     \pm 2.5   &                 &                  \\
6730.72  &                 &                 &                 &                 & 11.4    %%@
\pm 2.7   & 11.4    \pm 2.7   &                 &                  \\
7135.8   &                 &                 &                 &                 & 17.3    %%@
\pm 4.0   & 17.3    \pm 4.0   &                 &                  \\
9069     &                 &                 &                 &                 & 17.5    %%@
\pm 3.8   & 17.5    \pm 3.8   &                 &                  \\
E(B-V)   & 0.29    \pm 0.02  &                 &                 &                 & 0.15    %%@
\pm 0.03  &                 &                 &                  \\
\noalign{\smallskip}
{\rm Wavelength} & {\rm PN25}          & {\rm O300}            & {\rm B600}            & %%@
{\rm U900}            & {\rm PN6}             & {\rm O300}            & {\rm B600}            %%@
& {\rm U900}             \\
\noalign{\smallskip}
3727^{\mathrm a}     &         \pm 65    &                 &                 &         \pm %%@
65    & 21.3    \pm 15    &                 &                 & 21.3    \pm 15.2   \\
3868.76  & 152     \pm 38    &                 &                 & 152     \pm 38    & 127     %%@
\pm 31    &                 & 127     \pm 29    & 127     \pm 31     \\
4101.737 & 38      \pm 19    &                 &                 & 38      \pm 19    & 16.2    %%@
\pm 9.2   &                 &                 & 16.2    \pm 9.2    \\
4340.468 &         \pm 48    & 180     \pm 44    &                 &         \pm 48    & 47      %%@
\pm 15    &                 &         \pm 22    & 47      \pm 15     \\
4363.21  &         \pm 48    & 84      \pm 36    &                 &                 & 23      %%@
\pm 11    &                 &         \pm 22    & 23      \pm 11     \\
4685.682 &         \pm 38    &         \pm 73    &                 &         \pm 38    & 32      %%@
\pm 12    &         \pm 36    & 34      \pm 15    & 32      \pm 12     \\
4861.332 & 100     \pm 43    & 100     \pm 43    &         \pm 118   & 100     \pm 27    & %%@
100     \pm 20    & 100     \pm 20    & 100     \pm 17    &                  \\
4958.92  & 595     \pm 49    & 595     \pm 49    & 1115    \pm 547   &                 & 598     %%@
\pm 23    & 598     \pm 23    & 438     \pm 50    &                  \\
5006.85  & 1600    \pm 66    & 1600    \pm 66    & 3338    \pm 1626  &                 & %%@
1731    \pm 30    & 1731    \pm 30    & 1314    \pm 133   &                  \\
5875.666 & 35.9    \pm 4.1   & 35.9    \pm 4.1   &         \pm 102   &                 & %%@
29.6    \pm 6.5   &         \pm 26    & 29.6    \pm 6.5   &                  \\
6300.32  &         \pm 26    &         \pm 26    &                 &                 &         %%@
\pm 12    &         \pm 12    &                 &                  \\
6312.06  &         \pm 26    &         \pm 26    &                 &                 &         %%@
\pm 12    &         \pm 12    &                 &                  \\
6363.81  &         \pm 26    &         \pm 26    &                 &                 &         %%@
\pm 12    &         \pm 12    &                 &                  \\
6548.06  &         \pm 57    &         \pm 57    &         \pm 93    &                 &         %%@
\pm 15    &         \pm 15    &                 &                  \\
6562.817 & 285     \pm 32    & 285     \pm 32    & 470     \pm 236   &                 & 285     %%@
\pm 10    & 285     \pm 10    &                 &                  \\
6583.39  & 159     \pm 28    & 159     \pm 28    & 473     \pm 238   &                 & %%@
42.0    \pm 7.3   & 42.0    \pm 7.3   &                 &                  \\
7135.8   &         \pm 46    &         \pm 46    &                 &                 &         %%@
\pm 12    &         \pm 12    &                 &                  \\
9069     & 21.8    \pm 5.9   & 21.8    \pm 5.9   &                 &                 &                 %%@
&                 &                 &                  \\
E(B-V)   & 0.31    \pm 0.10  &                 &                 &                 & 0.17    %%@
\pm 0.03  &                 &                 &                  \\
\noalign{\smallskip}
\hline
\end{array}
\]
\begin{list}{}{}
\item[$^{\mathrm a}$] \lq\lq 3727" denotes the sum of
[\ion{O}{ii}]$\lambda\lambda$3726,3729
\end{list}
\end{table*}

\begin{table*}
\caption[]{Line Intensities for PNe in the bulge of \object{M31}}
\label{table5}
\[
\begin{array}{l*{7}r}
\hline
\noalign{\smallskip}
{\rm Wavelength} &     {\rm PN172}    &      {\rm PN31}      & {\rm PN80}       &    {\rm %%@
PN30}        &   {\rm PN29}          &  {\rm PN28} & {\rm PN23}  \\
\noalign{\smallskip}
3727     &         \pm 29    &         \pm 30    & 41.3    \pm 8.7   & 109     \pm 19    &         %%@
\pm 29    &         \pm 43    &         \pm 25   \\
3868.76  & 138     \pm 16    & 181     \pm 13    & 90.9    \pm 8.6   & 108     \pm 10    & %%@
104     \pm 15    & 95      \pm 16    & 137     \pm 21   \\
3967.47  & 56      \pm 10    & 59      \pm 11    & 38.7    \pm 8.4   & 44      \pm 10    & %%@
42      \pm 10    & 41      \pm 11    & 52      \pm 15   \\
4101.737 & 27.5    \pm 7.0   & 18.0    \pm 4.7   & 32      \pm 10    & 29.1    \pm 5.6   & %%@
25.9    \pm 6.4   & 21.8    \pm 9.0   & 25      \pm 12   \\
4340.468 & 47.0    \pm 5.4   & 47.0    \pm 4.5   & 47.0    \pm 5.9   & 47.0    \pm 6.3   & %%@
47.0    \pm 8.6   & 47.0    \pm 7.9   & 47.0    \pm 9.9  \\
4363.21  & 25.8    \pm 4.7   & 18.2    \pm 3.6   &         \pm 10    &         \pm 10    &         %%@
\pm 14    &         \pm 8     &         \pm 19   \\
4685.682 &         \pm 6     &         \pm 6     &         \pm 8     & 24.1    \pm 4.0   & %%@
21.6    \pm 5.0   &         \pm 10    &         \pm 14   \\
4861.332 & 100.0   \pm 5.0   & 100.0   \pm 4.4   & 100.0   \pm 5.0   & 100.0   \pm 4.9   & %%@
100.0   \pm 6.9   & 100.0   \pm 7.8   & 100.0   \pm 8.5  \\
4958.92  & 613     \pm 18    & 737     \pm 21    & 605     \pm 20    & 645     \pm 21    & %%@
691     \pm 30    & 685     \pm 32    & 459     \pm 26   \\
5006.85  & 1819    \pm 49    & 2253    \pm 59    & 1869    \pm 60    & 2063    \pm 63    & %%@
2219    \pm 91    & 2151    \pm 96    & 1347    \pm 75   \\
5875.666 & 15.6    \pm 2.3   &                 &                 &                 &                 %%@
&                 &                \\
E(B-V)   & 0.41    \pm 0.30  & 0.40    \pm 0.25  & -0.16   \pm 0.33  & 0.16    \pm 0.35  & -%%@
0.03   \pm 0.48  & 0.15    \pm 0.44  & 0.74    \pm 0.56 \\
\noalign{\smallskip}
{\rm Wavelength} & {\rm PN12} & {\rm PN10} & {\rm PN1} & {\rm PN3} & {\rm PN38} & {\rm PN36} %%@
& {\rm PN53} \\
\noalign{\smallskip}
3727     &         \pm 86    &         \pm 25    &         \pm 40    &         \pm 11    & %%@
18.9    \pm 7.0   &         \pm 31    & 22.6    \pm 6.2  \\
3868.76  & 176     \pm 30    & 70      \pm 12    & 158     \pm 18    & 64.6    \pm 6.9   & %%@
57.4    \pm 8.1   & 130     \pm 15    & 84.0    \pm 7.4  \\
3967.47  & 55      \pm 16    & 33      \pm 11    & 63      \pm 12    & 22.5    \pm 5.6   & %%@
23.9    \pm 6.8   & 103     \pm 15    & 37.7    \pm 6.6  \\
4101.737 & 26      \pm 11    &         \pm 19    & 31.0    \pm 8.6   &         \pm 13    & %%@
34.0    \pm 6.8   & 29.4    \pm 8.8   & 15.2    \pm 2.9  \\
4340.468 & 47.0    \pm 9.9   & 47      \pm 12    & 47.0    \pm 8.7   & 47.0    \pm 6.7   & %%@
47.0    \pm 5.9   & 47.0    \pm 5.4   & 47.0    \pm 4.5  \\
4363.21  & 24.7    \pm 8.0   &         \pm 18    &         \pm 12    &         \pm 12    &         %%@
\pm 9     & 20.7    \pm 4.5   & 11.8    \pm 3.4  \\
4685.682 & 21.2    \pm 7.7   &         \pm 15    &         \pm 15    &         \pm 11    &         %%@
\pm 8     &         \pm 7     &         \pm 5    \\
4861.332 & 100.0   \pm 8.6   & 100     \pm 11    & 100     \pm 10    & 100.0   \pm 9.1   & %%@
100.0   \pm 6.1   & 100.0   \pm 5.0   & 100.0   \pm 4.1  \\
4958.92  & 586     \pm 38    & 396     \pm 29    & 593     \pm 36    & 672     \pm 37    & %%@
408     \pm 16    & 382     \pm 13    & 674     \pm 18   \\
5006.85  & 1734    \pm 112   & 1183    \pm 81    & 1825    \pm 109   & 2153    \pm 115   & %%@
1256    \pm 44    & 1136    \pm 35    & 2062    \pm 52   \\
5875.666 & 10.6    \pm 3.6   & 25.3    \pm 7.1   &                 & 59.3    \pm 11.0  &                 %%@
& 8.8     \pm 1.6   & 38.2    \pm 5.3  \\
E(B-V)   & 0.88    \pm 0.56  & 0.18    \pm 0.68  & 0.49    \pm 0.49  & -0.55   \pm 0.38  & -%%@
0.09   \pm 0.33  & 0.78    \pm 0.30  & -0.31   \pm 0.25 \\
\noalign{\smallskip}
{\rm Wavelength} & {\rm PN52} & {\rm PN42} & {\rm PN45} & {\rm PN43} & {\rm PN48} & {\rm %%@
PN95} & {\rm PN47} \\
\noalign{\smallskip}
3727     &         \pm 63    &         \pm 19    & 26.1    \pm 7.7   &         \pm 14    &         %%@
\pm 10    & 19.7    \pm 6.1   & 19.5    \pm 7.5  \\
3868.76  & 63      \pm 12    & 122.2   \pm 8.0   & 120.5   \pm 7.8   & 56.7    \pm 6.9   & %%@
125.9   \pm 5.6   & 60.7    \pm 3.4   & 63.9    \pm 5.0  \\
3967.47  & 32      \pm 10    & 32.9    \pm 8.0   & 56.9    \pm 6.3   & 30.2    \pm 5.8   & %%@
42.7    \pm 5.8   & 28.5    \pm 4.8   & 35.4    \pm 6.6  \\
4101.737 & 43      \pm 14    & 16.8    \pm 4.1   & 25.2    \pm 5.4   & 14.8    \pm 4.3   & %%@
22.1    \pm 2.9   & 23.4    \pm 3.6   & 21.6    \pm 4.5  \\
4340.468 & 47.0    \pm 9.9   & 47.0    \pm 4.3   & 47.0    \pm 3.6   & 47.0    \pm 6.5   & %%@
47.0    \pm 3.3   & 47.0    \pm 3.4   & 47.0    \pm 4.4  \\
4363.21  &         \pm 13    & 25.9    \pm 3.7   & 27.2    \pm 3.2   &         \pm 11    & %%@
11.7    \pm 2.6   &         \pm 5     & 8.1     \pm 3.2  \\
4685.682 & 22.2    \pm 6.7   &         \pm 6     &         \pm 4     &         \pm 7     & %%@
4.0     \pm 1.9   &         \pm 3     &         \pm 4    \\
4861.332 & 100.0   \pm 8.2   & 100.0   \pm 4.7   & 100.0   \pm 3.4   & 100.0   \pm 4.9   & %%@
100.0   \pm 3.3   & 100.0   \pm 2.4   & 100.0   \pm 3.6  \\
4958.92  & 464     \pm 27    & 694     \pm 19    & 462     \pm 10    & 521     \pm 18    & %%@
700     \pm 14    & 331.1   \pm 5.8   & 474     \pm 10   \\
5006.85  & 1364    \pm 75    & 2111    \pm 54    & 1381    \pm 27    & 1576    \pm 52    & %%@
2160    \pm 43    & 998     \pm 15    & 1452    \pm 28   \\
5875.666 &         \pm 6     & 12.0    \pm 1.8   & 15.9    \pm 1.6   & 34.1    \pm 6.7   & %%@
36.1    \pm 3.7   & 18.6    \pm 1.5   & 20.9    \pm 2.4  \\
E(B-V)   & 0.63    \pm 0.56  & 0.30    \pm 0.24  & 0.44    \pm 0.20  & -0.72   \pm 0.36  & -%%@
0.25   \pm 0.19  & 0.11    \pm 0.19  & 0.09    \pm 0.25 \\
\noalign{\smallskip}
{\rm Wavelength} & {\rm PN408} & {\rm PN93} & {\rm PN92} & {\rm PN97} & {\rm PN91} & {\rm %%@
PN387} & {\rm PN380} \\
\noalign{\smallskip}
3727     &                 & 39.1    \pm 8.9   & 73      \pm 14    &         \pm 17    &         %%@
\pm 30    & 126     \pm 17    &         \pm 26   \\
3868.76  &                 & 84.2    \pm 6.8   & 129     \pm 13    & 79.5    \pm 8.2   & %%@
63.8    \pm 9.9   & 111     \pm 9.4   & 108     \pm 10   \\
3967.47  &                 & 27.3    \pm 5.4   & 44.8    \pm 8.5   & 44.5    \pm 4.3   & %%@
29.5    \pm 8.1   & 43.9    \pm 6.9   & 51.8    \pm 9.6  \\
4101.737 &                 & 21.9    \pm 4.8   & 34.7    \pm 7.9   & 21.7    \pm 4.5   &         %%@
\pm 14    & 30.3    \pm 6.2   & 28.8    \pm 6.8  \\
4340.468 &                 & 47.0    \pm 8.7   & 47.0    \pm 5.0   & 51.6    \pm 4.4   & %%@
47.0    \pm 6.0   & 47.0    \pm 3.5   & 50.4    \pm 5.7  \\
4363.21  &                 &         \pm 16    & 18.4    \pm 4.1   & 9.3     \pm 3.4   & %%@
18.9    \pm 4.8   & 12.3    \pm 2.7   & 30.5    \pm 5.1  \\
4685.682 &                 & 15.0    \pm 3.8   &         \pm 6     &         \pm 4     &         %%@
\pm 8     & 15.3    \pm 2.7   &         \pm 5    \\
4861.332 & 100     \pm 29    & 100.0   \pm 5.4   & 100.0   \pm 4.0   & 100.0   \pm 3.2   & %%@
100.0   \pm 5.3   & 100.0   \pm 4.0   & 100.0   \pm 4.6  \\
4958.92  & 491     \pm 85    & 673     \pm 24    & 565     \pm 15    & 565     \pm 13    & %%@
482     \pm 17    & 525     \pm 15    & 614     \pm 22   \\
5006.85  & 1431    \pm 242   & 2070    \pm 72    & 1703    \pm 42    & 1744    \pm 35    & %%@
1437    \pm 44    & 1591    \pm 36    & 1851    \pm 48   \\
5875.666 &                 & 30.0    \pm 6.0   & 17.2    \pm 1.8   & 22.7    \pm 2.4   &                 %%@
& 16.1    \pm 1.3   & 18.3    \pm 1.5  \\
6548.06  &                 &                 &                 & 12.0    \pm 2.1   &                 %%@
&                 & 4.4     \pm 1.8  \\
6562.817 &                 &                 &                 & 285.0   \pm 6.4   &                 %%@
&                 & 285.0   \pm 7.3  \\
6583.39  &                 &                 &                 & 35.8    \pm 2.5   &                 %%@
&                 & 16.9    \pm 2.0  \\
E(B-V)   &                 & -0.24   \pm 0.49  & 0.51    \pm 0.28  & 0.02    \pm 0.02  & %%@
0.26    \pm 0.34  & 0.39    \pm 0.20  & 0.20    \pm 0.02 \\
\noalign{\smallskip}
\hline
\end{array}
\]
\end{table*}


\end{document}